\documentclass[journal]{IEEEtran}

\usepackage{cite}
\usepackage[numbers,sort&compress]{natbib}
\usepackage{amsmath,amssymb,amsfonts}
\usepackage{graphicx}
\usepackage{subfigure}
\usepackage{textcomp}
\usepackage{xcolor}
\usepackage{transparent}
\usepackage{bm}
\usepackage{caption}
\usepackage{algpseudocode}

\usepackage{multirow}
\usepackage[table,xcdraw]{xcolor}
\usepackage{booktabs}

\usepackage{tabularx}

\usepackage{algorithm}

\usepackage{amsthm}

\usepackage{stfloats}
\usepackage{url}

\def\BibTeX{{\rm B\kern-.05em{\sc i\kern-.025em b}\kern-.08em
    T\kern-.1667em\lower.7ex\hbox{E}\kern-.125emX}}
\ifCLASSINFOpdf
\else
\fi

\begin{document}

\linespread {1.00}  
\addtolength{\parskip}{1.00pt} 

\setlength{\columnsep}{10.0pt}

% 公式间距
\makeatletter
\renewcommand\normalsize{%
\@setfontsize\normalsize\@xpt\@xiipt
\abovedisplayskip 4\p@ \@plus2\p@ \@minus5\p@
\abovedisplayshortskip \z@ \@plus3\p@
\belowdisplayshortskip 6\p@ \@plus3\p@ \@minus3\p@
\belowdisplayskip \abovedisplayskip
\let\@listi\@listI}
\makeatother

\title{WiFo-MUD: Wireless Foundation Model for Heterogeneous Multi-User Demodulator}

\author{Zonghui~Yang,~\IEEEmembership{Graduate~Student~Member,~IEEE},
Shijian~Gao,~\IEEEmembership{Member,~IEEE},\\
Xuesong~Cai,~\IEEEmembership{Senior Member,~IEEE},
Xiang~Cheng,~\IEEEmembership{Fellow,~IEEE},
Liuqing~Yang,~\IEEEmembership{Fellow,~IEEE}.

\thanks{Z.~Yang, X.~Cai and X.~Cheng are with the State Key Laboratory of Photonics and Communications, School of Electronics, Peking University, Beijing 100871, China (e-mail: yzh22@stu.pku.edu.cn; xuesong.cai@pku.edu.cn; xiangcheng@pku.edu.cn).}% <-this % stops a space
\thanks{S.~Gao is with the Internet of Things Thrust, The Hong Kong University of Science and Technology (Guangzhou), Guangzhou 511400, China (e-mail: shijiangao@hkust-gz.edu.cn).}% <-this % stops a space
\thanks{L.~Yang is with the Internet of Things Thrust \& Intelligent Transportation Thrust, The Hong Kong University of Science and Technology (Guangzhou) Guangzhou, China, and also with the Department of Electronic and Computer Engineering, Hong Kong University of Science and Technology, Hong Kong SAR, China (Email: lqyang@ust.hk).}
}
\maketitle

\markboth{IEEE TRANSACTIONS, 2025} %
{Shell \MakeLowercase{\textit{et al.}}: Bare Demo of IEEEtran.cls for IEEE Journals}

\begin{abstract}

Multi-user signal demodulation is critical to wireless communications, directly impacting transmission reliability and efficiency. However, existing demodulators underperform in generic multi-user environments: classical demodulators struggle to balance accuracy and complexity, while deep learning-based methods lack adaptability under heterogeneous configurations. Although diffusion models have been introduced for demodulation, their flexibility remains limited for practical use. To address these issues, this work proposes WiFo-MUD, a universal diffusion-based foundation model for multi-user demodulation. The model aligns inter-user signal-to-noise ratio imbalance and performs conditional denoising via a customized backbone. Furthermore, a communication-aware consistency distillation method and a dynamic user-grouping strategy are devised to enhance inference. WiFo-MUD achieves state-of-the-art results on large-scale heterogeneous datasets, demonstrating efficient inference and strong generalization across varying system configurations.

\end{abstract}

\begin{IEEEkeywords}
Wireless foundation model, multi-user demodulation, diffusion model, consistency distillation.
\end{IEEEkeywords}

\IEEEpeerreviewmaketitle

\vspace{-0.1cm}
\section{Introduction}

\IEEEPARstart{M}ULTI-user signal demodulation constitutes a critical task in the design of digital communication systems \cite{overview_MIMO_detection, overview_massive_MIMO_detection, 3GPP_R18}. Accurate signal demodulation ensures reliability of data transmission, and reduces retransmission rates, thereby enhancing overall spectral efficiency \cite{SIMOFDM, BPM}. As such, it underpins the performance and viability of emerging applications, including autonomous driving, low-altitude economy and embodied intelligence \cite{overview_Som}.

The demodulation problem is known to be NP-hard owing to the discrete nature of transmitted symbols.
Although maximum likelihood (ML) detection theoretically has the optimal performance, it has a complexity increasing exponentially as the number of transmitting antennas and modulation order, which is prohibitive for practical systems.
Linear demodulators, such as least squares (LS) and linear minimum mean square error (LMMSE) \cite{LS}, are a class of methods where the received signal is simply multiplied by a filter matrix to obtain an estimate and then make the decision. While linear methods are favored for their low complexity, they incur substantial performance degradation in large-scale multiple-input multiple-output (MIMO) systems and with higher-order modulations.
Compared to linear demodulators, nonlinear demodulation has a better performance with increased complexity \cite{sphere_decoding, MP_Detector, AMP, OAMP}. For instance, the sphere decoding algorithm reduces the search space of ML detection by some restrictions. Message passing-based methods \cite{MP_Detector, AMP} achieve near-optimal performance under ideal Gaussian channels but is notably sensitive to deviations from this assumption. Orthogonal approximate message passing (OAMP) \cite{OAMP} was later proposed to relax the constraints on channel characteristics.
However, above-mentioned demodulators lack a trade-off between computational complexity and demodulation accuracy: Linear demodulators are efficient but suboptimal, while non-linear methods approach optimal performance at a high computational cost.

%The semi-definite relaxation (SDR) algorithm proposed transforms the maximum likelihood problem into a quadratic programming problem, and then relaxes the constraints to become a semi definite programming problem, which can be solved by interior point method. The advantage is that there is no local minimum according to convex optimization theory, and for the transmitting antenna, it has polynomial complexity in modulation order.

The emergence of deep learning has advanced the development of data-driven designs for physical-layer transmission \cite{TCOM, TWC}. Neural networks, capable of learning complex mappings from channel distributions and received signals to transmitted symbols, offer promising tools for signal demodulation and achieve better balance. \cite{model_driven_MIMO_detection, dl_detector_TWC}
Models like DnCNN \cite{CNN_detector} directly leveraged convolutional neural networks (CNN) for signal demodulation but the accuracy is limited. 
Other approaches, such as RE-MIMO \cite{RE-MIMO} and OAMP-Net \cite{OAMPNet}, integrated model-based iterations with neural networks to enhance performance. Nevertheless, they still struggle to generalize to correlated multi-user channels. 
Furthermore, all these learning-based models exhibit poor generalization across heterogeneous system configurations and shifting data distributions. Their performance markedly degrades under varying antennas, modulations or out-of-distribution channel conditions, highlighting a critical gap in robustness.

In recent years, the powerful generative and modeling capabilities demonstrated by diffusion models in domains such as computer vision and natural language processing have motivated their adoption in wireless communications \cite{DDPM, SDE}. These models offer potential for characterizing complex channel and noise distributions, which could enhance signal demodulation accuracy \cite{ALD_2023TWC, Joint_ALD_ICASSP, PC_sampler, CoDiPhy, CDDM}.
%Initial work by \cite{SNIPS} introduced diffusion models for solving noisy linear inverse problems. Inspired by this, 
For instance, \cite{ALD_2023TWC} developed an annealed Langevin dynamics (ALD)-based demodulation algorithm, which uses annealing to facilitate the training of a channel-dependent score network and iteratively refines signal estimates. 
Furthermore, \cite{Joint_ALD_ICASSP} proposed a joint channel and signal estimation method based on diffusion models, and \cite{PC_sampler} improved demodulation accuracy via a predictor-corrector scheme.  
Studies in \cite{CDDM} utilized diffusion models for end-to-end robust image semantic transmission, yet these methods fail to incorporate realistic channel distributions.

%In a different vein, [2025WCL] exploited the dual denoising and generative capabilities of diffusion models to enhance channel estimation and generate synthetic channel data. This method adapted the number of denoising steps according to SNR fluctuations among received signals, thereby accounting for channel variability.

Although diffusion-based methods provide novel perspectives for signal demodulation, significant challenges impede their practical deployment.
First, these approaches overlook realistic multi-user scenarios, where performance may be considerably degraded by inter-user interference (IUI) and imbalanced signal-to-noise ratios (SNR) across users. 
Furthermore, they demonstrate limited flexibility in adapting to dynamic changes, such as variations in user count or channel dimensions, and fail to adequately incorporate multi-user channel correlations. Consequently, existing methods lack generalizability across heterogeneous multi-user environments, resulting in compromised demodulation accuracy.
Recent studies have explored wireless foundation models \cite{WiFo, WiFo_CF, overview_FM_TNSE}, enhancing performance while maintaining adaptability across heterogeneous systems. However, none of these existing models is capable of addressing this specific signal demodulation task.

%First, despite attempts to accelerate sampling, these approaches still demand substantial iterations to achieve satisfactory solutions, rendering them unsuitable for low-latency, real-time communication systems. However, this method requires iterations and involves computationally expensive operations such as singular value decomposition (SVD). To alleviate computational burden, \cite{DM_detector_huangyongming} proposed an approximate diffusion detection as a more flexible and SVD-free alternative, and \cite{Accelerate_ALD_ICASSP} incorporated annealed underdamped Langevin dynamics, introducing a momentum term to accelerate sampling.

To address the aforementioned limitations, this paper proposes WiFo-MUD, a foundation model-empowered framework tailored for multi-user signal demodulation under heterogeneous system configurations. The framework first mitigates multi-user SNR gaps using dedicated alignment modules, followed by a denoising process applied to coarse estimates via a wireless diffusion Transformer architecture, which incorporates structural priors of wireless signals. Leveraging a diffusion-based pre-training strategy, WiFo-MUD effectively captures diverse channel and noise distributions, leading to improved demodulation accuracy over existing methods. Capitalizing on the flexibility of the Transformer backbone and customized embedding modules, the model also supports dynamic adaptation to varying numbers of users and antennas.
Nevertheless, the iterative reverse process in diffusion models can introduce high inference latency. While prior efforts such as \cite{DM_detector_huangyongming, Accelerate_ALD_ICASSP} have introduced approximation techniques or momentum terms to reduce complexity, they still require considerable iterations to achieve satisfactory performance. To overcome this bottleneck, we further devise a communication-aware consistency distillation scheme that eliminates iterative refinement, substantially cutting down latency and fulfilling real-time demands. Additionally, a user-grouping strategy is incorporated to mitigate inter-user interference. When pre-trained and distilled on a large-scale heterogeneous signal demodulation dataset, WiFo-MUD demonstrates superior performance in both full-shot and zero-shot scenarios, enabling reliable and efficient communication in multi-user systems with diverse configurations.

\begin{figure*}[t]
  \vspace{-0.2cm}
  \centering
  \includegraphics[width=0.75\linewidth]{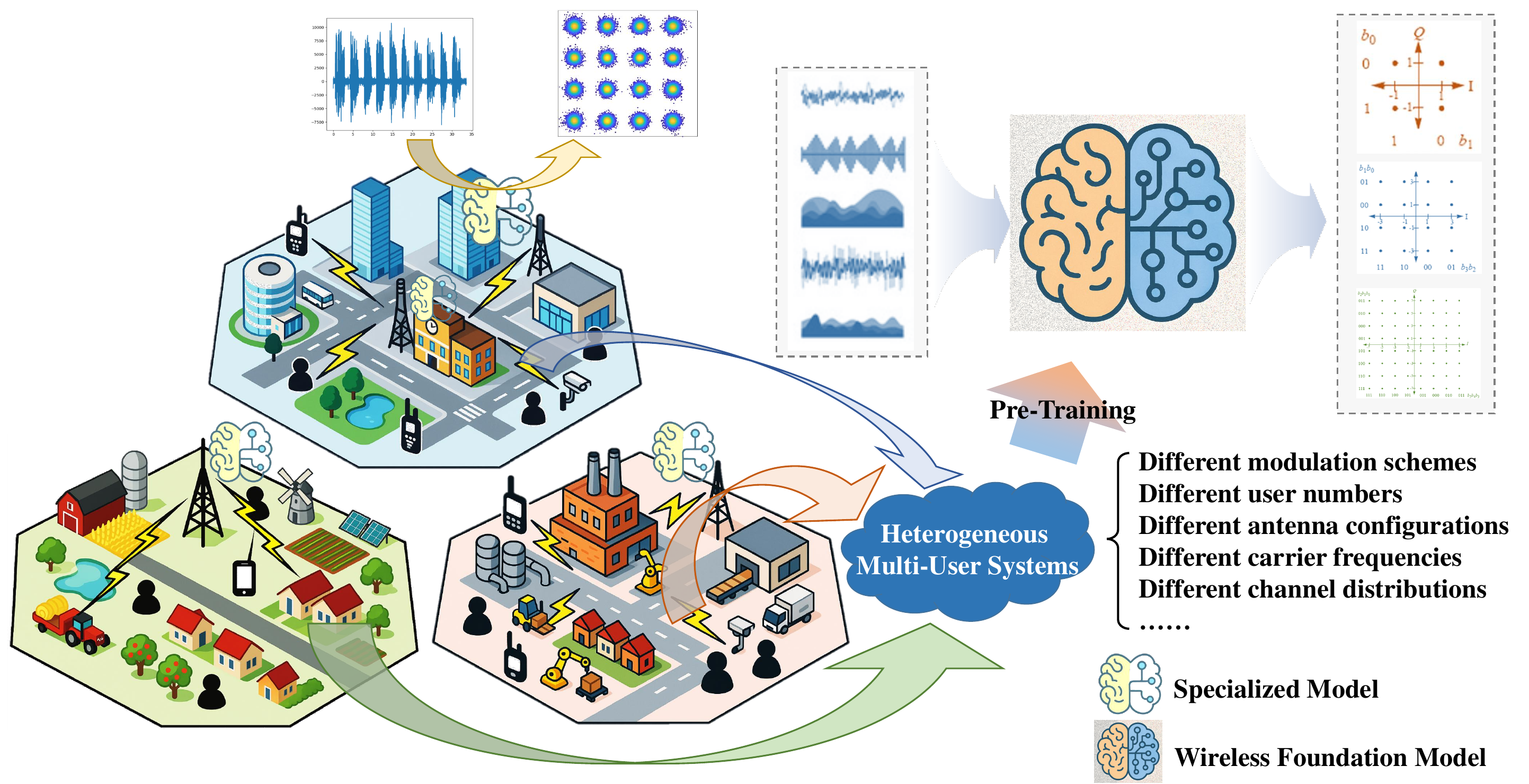}
  \vspace{-0.0cm}
  \captionsetup{font=small}
  \caption{Comparison of model paradigms for heterogeneous multi-user signal demodulation.}
  \label{fig:system_model}
  \vspace{-0.3cm}
\end{figure*}

Our contributions can be summarized as follows:
\vspace{-0.1cm}
\begin{itemize}
    \item We propose WiFo-MUD, a foundation model for heterogeneous multi-user signal demodulation that integrates a multi-user alignment module with a wireless diffusion Transformer backbone. Through diffusion-based pre-training, the model achieves high accuracy and strong generalization across varying system configurations.
    \item To enhance inference efficiency, we design a communication-aware consistency distillation technique for accurate single-step inference and a dynamic user-grouping mechanism to mitigate inter-user interference.
    \item Extensive experiments on a large-scale heterogeneous dataset demonstrate that our model surpasses existing benchmarks in both full-shot and zero-shot scenarios, achieving low-latency inference while maintaining robust generalization.
\end{itemize}

\newcommand{\RNum}[1]{\uppercase\expandafter{\romannumeral #1\relax}}

The remainder of this work is structured as follows. Section \RNum{2} introduces the system model and the problem formulation. Section \RNum{3} includes a brief background of diffusion models as a preliminary. The wireless foundation model for multi-user signal demodulation is illustrated in Section \RNum{4}, and further optimization for model inference is introduced in Section \RNum{5}. Section \RNum{6} contains our simulation findings, and Section \RNum{7} concludes our work.

\textit{Notation}: $a$, $\bm a$ and $\bm A$ represent a scalar, a vector and a matrix respectively. $(\cdot)^{\text{T}}$, $(\cdot)^{\text{H}}$, $\text{Tr}(\cdot)$, $\text{rank}(\cdot)$, $\left\|\cdot\right\|_2$ and $\left\|\cdot\right\|_{\text{\text{F}}}$ denote transpose, conjugate transpose, trace, rank, 2-norm and Frobenius norm, respectively. $\mathbb{E}(\cdot)$ denotes the expectation. $\mathbb{P}(\cdot)$ denotes the probability. $\mathcal{CN}(m,\sigma^2)$ represents the complex Gaussian distribution whose mean is $m$ and covariance is $\sigma^2$. $\mathbb{R}$ and $\mathbb{C}$ denote the set of real numbers and complex numbers respectively. $\vert z \vert$ and $\angle z$ denote the modulus and the phase of a complex number $z$ respectively. $\bigcap$ and $\bigcup$ denote the intersection and the union, and $\emptyset$ denotes an empty set.

%\newcommand{\RNum}[1]{\uppercase\expandafter{\romannumeral #1\relax}}

%%%%%%%%%%%% System Model and Problem Formulation %%%%%%%%%%%

\section{System Model and Problem Formulation}

\subsection{System Setup}

We consider $U$ active users in one cell, the $u$-th of which having $N_{t_{u}}$ antennas, transmitting signals simultaneously to the base station (BS) equipped with $N_{r}$ antennas. The signal received at the BS is given by
\vspace{-0.2cm}
\begin{equation}
    \bm y=\sum_{u=1}^{U}\bm H_{u}\bm x_{u} + \bm n,
    \vspace{-0.1cm}
\end{equation}
where $\bm H_{u} \in \mathbb{C}^{N_{r} \times N_{t_u}}$ denotes the uplink channel matrix from the $u$-th user to the BS, and $\bm n \sim \mathcal{CN}(\bm 0, \sigma_n^2 \bm I)$ represents a complex circularly symmetric Gaussian noise vector with variance $\sigma_n^2$. The vector $\bm x_u \in \mathcal{X}^{N_{t_u}}$ contains the transmitted symbols from the $u$-th user, where $\mathcal{X}$ is a finite set of constellation points. Various modulation schemes may be employed, and the transmitted symbols are normalized to have unit average power per antenna, i.e. $\mathbb{E}\left[ \|\bm x_{u}\|_{2}^{2}\right ]=N_{t_{u}}$. The same constellation set is assumed for all users, and each symbol has the same probability of being chosen by the users. Furthermore, statistical characteristics of received noise $\sigma_{n}$ is assumed to be known, and slot-level channel state information (CSI) is assumed to be estimated at BS, denoted by $\hat{\bm H}_{u}$. The mean squared error of channel estimation is expressed as $\sigma_{H,u}^{2}=\mathbb{E}\left [\|\hat{\bm H}_{u}-\bm H_{u}\|_{\text{F}}^{2}\right ]$. We use the estimated channels and the received signal to recover the transmitted symbols.

\subsection{Heterogeneous Multi-User Signal Demodulation}

Given the observed channel $\hat{\bm H}_{u}$ for $u = 1, \cdots, U$ and the observed signal $\bm y$ at the BS, the objective is to recover $\bm x$ using a demodulator $\mathcal{D}$ that maximizes the posterior probability. That is,
\vspace{-0.2cm}
\begin{equation}
\begin{aligned}
    \hat{\bm x}_{\text{MAP}} &= \mathcal{D}(\bm y, \{\hat{\bm H}_{u}\}_{u=1}^{U})  \\
    & = \underset{ \{\bm x_{u}\}_{u=1}^{U}\in \mathcal{X}}{\text{argmax}}~    p(\bm x \vert \bm y, \{\hat{\bm H}_{u}\}_{u=1}^{U}).
\end{aligned}
\label{equ:MAP}
\end{equation}
Under the assumption of a uniform prior distribution over the constellation points and Gaussian measurement noise, this maximum a posteriori (MAP) demodulator simplifies to a ML demodulator. Specifically, the estimation problem in Eq. (\ref{equ:MAP}) reduces to solving the following optimization problem as
\vspace{-0.2cm}
\begin{equation}
\begin{aligned}
    \hat{\bm x}_{\text{ML}} =\underset{\{\bm x_{u}\}_{u=1}^{U}\in \mathcal{X}}{\text{argmin}}~ \|\bm y-\sum_{u=1}^{U}\hat{\bm H}_{u}\bm x_{u}\|_{2}^{2},
\end{aligned}
\end{equation}
subject to the constraint that each $\bm x_u$ lies in a finite discrete constellation. 
This problem is known to be NP-hard, rendering the exact computation of $\hat{\bm x}_{\text{ML}}$ intractable in practical systems, where its combinatorial complexity fundamentally conflicts with strict latency and hardware constraints, especially as the number of users and modulation order increase.
Existing methods have sought to address this issue with manageable computational complexity, yet they struggle to achieve a satisfactory balance between accuracy and latency. Furthermore, many approaches fail to adequately account for the impact of noisy observations and heterogeneous configurations, such as varying $U$, $N_{t_{u}}$, $N_{r}$ and $\mathcal{X}$.

%%%%%%%%%%%% Diffusion Model preliminary %%%%%%%%%%%%%%%%%%%
\section{Background on Diffusion Models}
\label{sec:background_diffusion}

To start with, we briefly introduce the background of diffusion model as the preliminary, based on which we construct our framework.
Diffusion models \cite{DDPM, SDE} are a class of generative models that learn data distributions by inverting a gradual noising process. A unified perspective on these models is provided through stochastic differential equations (SDEs), which describe the continuous-time dynamics of diffusion.

\vspace{-0.2cm}
\subsection{SDE Formulation of Diffusion Processes}

The forward diffusion process can be expressed as an Ito SDE of the form:
\vspace{-0.1cm}
\begin{equation}
d\bm x = \bm f(\bm x, t) dt + g(t) d\bm w,
\label{equ:diffusion_contin}
\end{equation}
where $\bm w$ is a standard Wiener process, $\bm f(\bm x, t)$ is a drift coefficient, and $g(t)$ is a diffusion coefficient. This formulation generalizes various discrete-time diffusion processes.

The reverse-time generative process corresponding to Eq.~(\ref{equ:diffusion_contin}) is given by the reverse SDE as
\begin{equation}
d\bm x = \left[\bm f(\bm x, t) - g(t)^2 \nabla_{\bm x} \log p_t(\bm x)\right] dt + g(t) d\bar{\bm w},
\end{equation}
where $\bar{\bm w}$ is a Wiener process operating backwards in time, and $\nabla_{\bm x} \log p_t(\bm x)$ is the score function of the perturbed data distribution at time $t$.

Beyond the stochastic reverse-time SDE, diffusion models also admit an associated deterministic process described by an ordinary differential equation (ODE), known as the probability flow ODE \cite{SDE}. This ODE is derived by removing the Brownian noise term from the reverse SDE, yielding
\begin{equation}
d\bm x = \left[\bm f(\bm x, t) - \frac{1}{2}g(t)^2 \nabla_{\bm x} \log p_t(\bm x)\right] dt.
\end{equation}
Notably, this ODE defines a deterministic trajectory that shares the same marginal probability densities ${p_t(\bm x)}_{t=0}^T$ as the forward diffusion SDE. This formulation enables exact likelihood computation and more efficient sampling through adaptive ODE solvers, which would be implemented in the inference acceleration.

\begin{figure*}[t]
  \vspace{-0.2cm}
  \centering
  \includegraphics[width=0.85\linewidth]{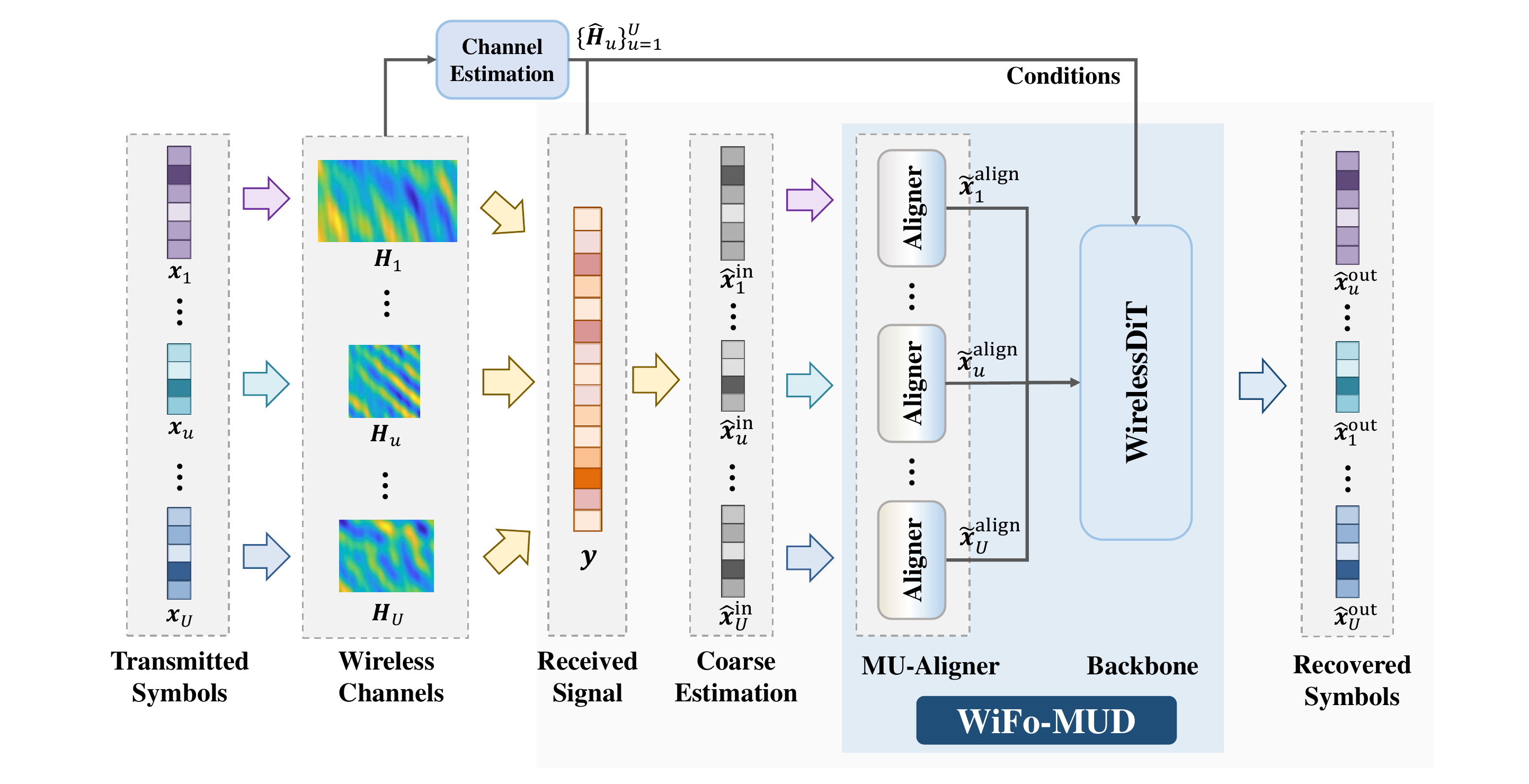}
  \vspace{-0.2cm}
  \captionsetup{font=small}
  \caption{The workflow of proposed WiFo-MUD in heterogeneous multi-user systems.}
  \label{fig:wifo_det}
  \vspace{-0.45cm}
\end{figure*}

%For both the VP-SDE (DDPM) and VE-SDE (score-based) formulations, the corresponding probability flow ODE can be discretized and simulated using a learned score function $\nabla_{\mathbf{x}} \log p_t(\mathbf{x})$ or noise predictor $\boldsymbol{\epsilon}_\theta(\mathbf{x}_t, t)$. This deterministic interpretation bridges diffusion models with neural ODEs and enhances their practicality by enabling faster sampling and improved control over the generative process.

\vspace{-0.2cm}
\subsection{Discretization and Specific Instances}

By choosing specific forms for $\bm f(\bm x, t)$ and $g(t)$, and discretizing the SDE in steps, we could recover the popular diffusion frameworks.

Setting $\bm f(\bm x, t) = -\frac{1}{2}\beta(t)\bm x$ and $g(t) = \sqrt{\beta(t)}$ with a linear noise schedule $\beta(t)$ yields the variance preserving (VP)-SDE. Discretizing this with $\beta_t = \beta(t)\Delta t$ leads to the denoising diffusion probabilistic model (DDPM) \cite{DDPM}. The forward process becomes $\bm x_{t} = \sqrt{1 - \beta_t} \bm x_{t-1} + \sqrt{\beta_{t}} \bm \epsilon, \quad \bm \epsilon \sim \mathcal{N}(\mathbf{0}, \mathbf{I})$.

%with the closed-form expression for any timestep:
%\begin{equation}
%\mathbf{x}_t = \sqrt{\bar{\alpha}_t} \mathbf{x}_0 + \sqrt{1 - \bar{\alpha}_t} \boldsymbol{\epsilon},
%\end{equation}
%where $\alpha_t = 1 - \beta_t$ and $\bar{\alpha}t = \prod{s=1}^t \alpha_s$.

Alternatively, setting $\bm f(\bm x, t) = \bm 0$ and $g(t) = \sqrt{\frac{d\sigma^2(t)}{dt}}$ with $\sigma(t)$ increasing yields the variance exploding (VE)-SDE, underlying score-based models \cite{SDE}. The forward process adds noise with variance $\sigma^2(t)$ as $\bm x_t = \bm x_0 + \sigma(t) \bm \epsilon$.

In both cases, the model learns to approximate the score function $\nabla_{\bm x} \log p_t(\bm{x})$ or equivalently, for denoising diffusion probabilistic models (DDPM), to predict the noise $\bm {\epsilon}_\theta(\bm x_t, t)$ via the objective as
\vspace{-0.1cm}
\begin{equation}
\mathcal{L} = \mathbb{E}_{t, \bm {x}_0, \bm{\epsilon}} \left[ \lambda(t) \| \bm{\epsilon} - \bm{\epsilon}_{\theta}(\mathbf{x}_t, t ) \|_{2}^2 \right],
\vspace{-0.1cm}
\end{equation}
where $\lambda(t)$ is a weighting function. This SDE perspective provides a unified framework for understanding and developing diffusion models, facilitating both theoretical analysis and practical algorithm design.

\vspace{-0.0cm}
\section{Wireless Foundation Model for Multi-User Signal Demodulation}

To overcome the bottlenecks of existing demodulators, based on diffusion models, we propose a wireless foundation model for signal demodulation, termed as \textbf{WiFo-MUD}, a flexible and scalable model for multi-user signal demodulation under heterogeneous system configurations. Integrating a multi-user alignment and a conditional denoising diffusion Transformer backbone, WiFo-MUD aims to achieve strong generalization across varying system configurations, while maintaining high accuracy and addressing the issue of multi-user imbalance.

Specifically, a coarse estimation of $\{\hat{\bm x}^{\text{in}}_{u}\}_{u=1}^{U}$ is calculated rapidly, and a \textbf{multi-user aligner (MU-Aligner)} pre-processes these estimates according to varying SNRs across users. Then these estimates are fed into a \textbf{wireless diffusion Transformer (WirelessDiT)} for denoising to refine the estimate as $\hat{\bm x}_{u}^{\text{out}}$. The workflow of WiFo-MUD is illustrated in Fig.~\ref{fig:wifo_det}.

\vspace{-0.2cm}
\subsection{Multi-User Aligner}

Directly incorporating coarse symbol estimates of different users into the model may result in performance degradation, due to high noise levels and significant SNR imbalance among users. 
To alleviate this, we first introduce a light-weight \textbf{MU-Aligner} as a pre-processing stage. In this module, the equivalent noise power of each coarse estimate is computed to determine an equivalent denoising timestep. Based on this, the MU-Aligner performs user-specific noise suppression and SNR alignment through a diffusion-based denoising mechanism.

\subsubsection{Timestep Calculation}

The coarse symbol estimate for the $u$-th user, denoted as $\hat{\bm x}_{u}$, can be obtained using low-complexity linear methods. For instance, the LMMSE estimate is given by
\vspace{-0.2cm}
\begin{equation}
    \hat{\bm x}_{u}=(\hat{\bm H}^{\text{H}}\hat{\bm H} + \sigma_{n}^{2}\bm R_{u}^{-1})^{-1}\hat{\bm H}^{\text{H}}\bm y,
    \vspace{-0.1cm}
\end{equation}
with $\bm R_{u}=\mathbb{E}[\bm x_{u}\bm x_{u}^{\text{H}}]=\bm I_{N_{t_{u}}}$. Defining the estimation error as $\bm \xi_{u}=\hat{\bm x}_{u}-\bm x_{u}$, the conditional error covariance given $\bm H_{u}$ is $\mathbb{E}[\bm \xi_{u}\bm \xi^{\text{H}}_{u}\vert \bm H_{u}]=\bm A_{u}^{-1}\left[(\sigma_{n}^{2}+\sigma_{H,u}^{2}N_{t_{u}})\bm H_{u}^{\text{H}}\bm H_{u}+ \sigma_{n}^{4}\bm I \right]\bm A_{u}^{-1}$, 
with $\bm A_{u}=\bm H_{u}^{\text{H}}\bm H_{u}+\sigma_{n}^{2}\bm R_{x}^{-1}=\bm H_{u}^{\text{H}}\bm H_{u}+\sigma_{n}^{2}\bm I_{N_{t_u}}$.

To estimate the error variance under current channel conditions, access to channel statistics is required. However, such information is often difficult to obtain in practical systems. Instead, we leverage historically measured CSI stored in memory buffers as a sampling of the full channel distribution. With sufficient memory samples, this collection asymptotically approximates the long-term channel statistics. Based on this statistics, we marginalize out $\bm H_{u}$, and the equivalent noise power can be calculated as 
%$\bar{\sigma}_{u}^{2}=\mathbb{E}_{\bm H_{u}}\left [\|\bm \xi_{u} \|_{2}^{2}\right]= \mathbb{E}_{\bm H_{u}} \left [\text{Tr}\left(\mathbb{E}[\bm \xi_{u}\bm \xi^{\text{H}}_{u}\vert \bm H_{u}] \right ) \right ] =(\sigma_{n}^{2}+\sigma_{H,u}^{2}N_{t_{u}})\text{Tr}(\mathbb{E}_{\bm H_{u}}\left [\bm H_{u}^{\text{H}}\bm H_{u}(\bm A_{u})^{-2}\right ]) +\sigma_{n}^{4}\text{Tr}(\mathbb{E}_{\bm H_{u}}\left [\bm A_{u}^{-2}\right ])$.
\vspace{-0.1cm}
\begin{align}
&~~~\bar{\sigma}_{u}^{2}=\mathbb{E}_{\bm H_{u}}\left [\|\bm \xi_{u} \|_{2}^{2}\right]=
    \mathbb{E}_{\bm H_{u}} \left [\text{Tr}\left(\mathbb{E}[\bm \xi_{u}\bm \xi^{\text{H}}_{u}\vert \bm H_{u}] \right ) \right ]  \\
    & \!\!=\!\!(\!\sigma_{n}^{2}\!\!+\!\sigma_{H,u}^{2}\!N_{t_{u}}\!)\text{Tr}\left(\mathbb{E}_{\bm H_{u}}\!\!\left [\bm H_{u}^{\text{H}}\bm H_{u}(\!\bm A_{u})^{-2}\right ]\!\right)\! +\!\sigma_{n}^{4}\text{Tr}\left(\mathbb{E}_{\bm H_{u}}\!\!\left [ \!\bm A_{u}^{-2}\right ]\!\right)\!.  \nonumber
\end{align}
In the $T$-step diffusion process, each denoising timestep corresponds to a specific level of equivalent noise power. Following the assumption in \cite{dual_use} that the difference in noise power between two adjacent steps is constant, the source timestep corresponding to the estimate $\hat{\bm x}_{u}$ is determined by the following relation:
\vspace{-0.2cm}
\begin{equation}
    \hat{t}_{u} = \frac{\bar{\sigma}_{u}^{2}(T-1)}{\bar{\alpha}_{1} - \bar{\alpha}_{T}}.
\end{equation}

\subsubsection{User-specific Denoising}
\begin{figure*}[t]
  \vspace{-0.1cm}
  \centering
  \includegraphics[width=0.8\linewidth]{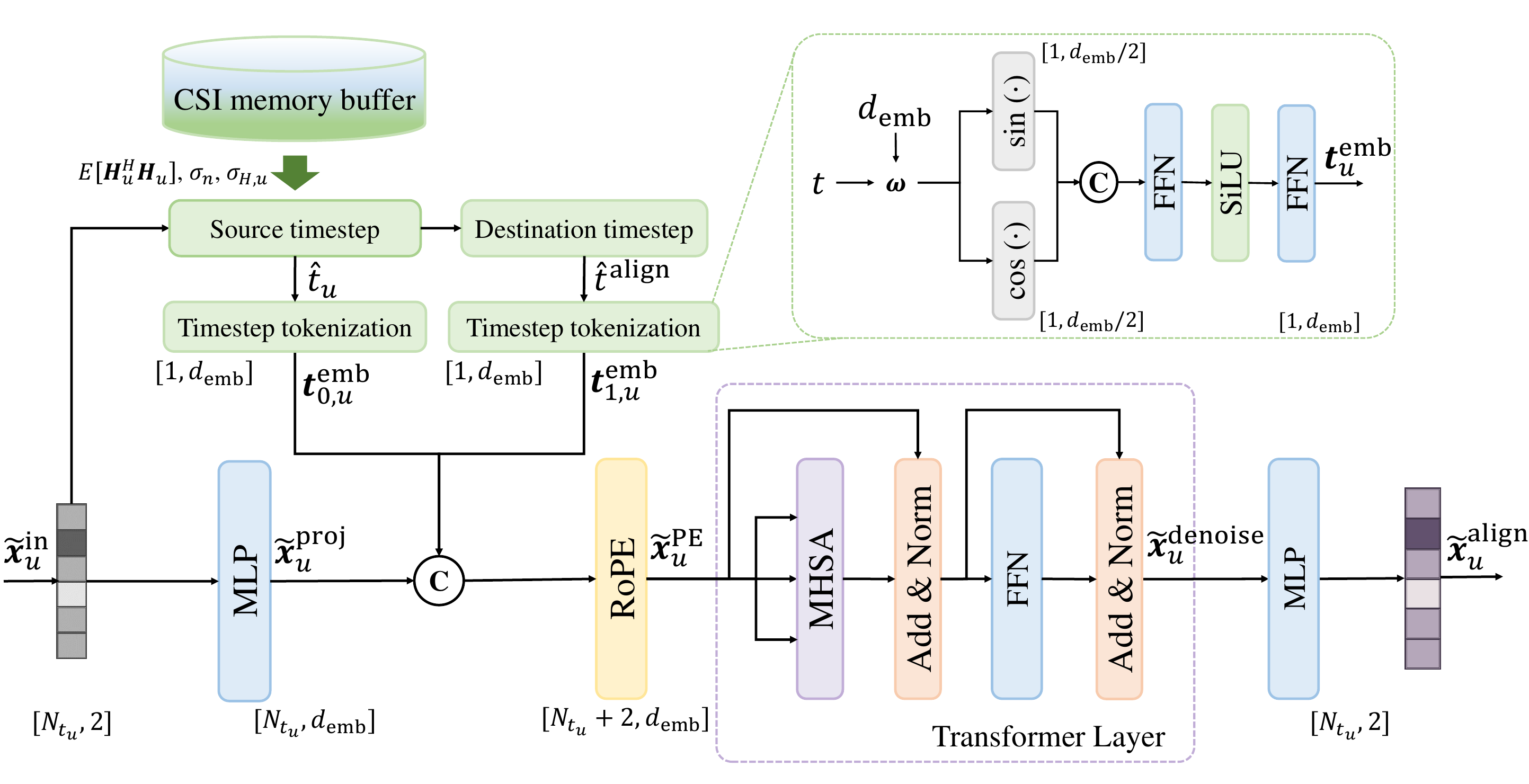}
  \vspace{-0.15cm}
  \captionsetup{font=small}
  \caption{The network architecture of MU-Aligner module in WiFo-MUD.}
  \label{fig:mu_aligner}
  \vspace{-0.4cm}
\end{figure*}

With the source timestep of the $u$-th user $\hat{t}_{u}$, the remaining task is to denoise $\hat{\bm x}_{u}$ from $\hat{t}_{u}$ to a destination timestep $\hat{t}^{\text{align}}=\text{argmin}_{u}\hat{t}_{u}$. The detailed architecture of the MU-Aligner is demonstrated in Fig.~\ref{fig:mu_aligner}.

\textbf{Input projection}:
Denote $\tilde{\bm x}^{\text{in}}_{u}=\left[\text{Re}\{\hat{\bm x}^{\text{in}}_{u}\}, \text{Im}\{\hat{\bm x}^{\text{in}}_{u}\} \right]\in \mathbb{R}^{N_{t_u}\times 2}$ as the real-valued input of the $u$-th estimate. $\tilde{\bm x}^{\text{in}}_{u}$ is first projected into a high-dimensional latent space as $\tilde{\bm x}_{u}^{\text{proj}}=\text{FFN}(\tilde{\bm x}_{u}^{\text{in}})\in \mathbb{R}^{N_{t_u}\times d_{\text{emb}}}$,
to obtain $N_{t_{u}}$ signal tokens, where $\text{FFN}(\cdot)$ represents one layer of the feedforward neural network (FFN) and $d_{\text{emb}}$ is the dimension of the latent space.

\textbf{Position embedding}:
To capture sequential dependencies within the input, relative positional information is encoded using rotary position embedding (RoPE), yielding $\tilde{\bm x}_{u}^{\text{PE}} = \text{RoPE}(\tilde{\bm x}_{u}^{\text{proj}})$. This method adapts to variable input lengths by embedding positions through rotations in the complex plane, and introduces a distance-related decay in attention scores while preserving directional relationships between tokens.

%The embedded tokens can be written as
%\vspace{-0.1cm}
%\begin{align}
%    \tilde{\bm x}_{u}^{\text{PE}}[n, 2j]&\!=\!\cos(n\theta_{j}) \tilde{\bm x}_{u}^{\text{proj}}[n, 2j]\!-\!\sin(n\theta_{j})\tilde{\bm x}_{u}^{\text{proj}}[n, 2j\!+\!1], \\
%    \tilde{\bm x}_{u}^{\text{PE}}[n, 2j\!+\!1]&\!=\!\sin(n\theta_{j}) \tilde{\bm x}_{u}^{\text{proj}}[n, 2j]\!+\!\cos(n\theta_{j})\tilde{\bm x}_{u}^{\text{proj}}[n, 2j\!+\!1],
%\end{align}
%where $\theta_{j}=10000^{-2j/d_{\text{emb}}}$ is the frequency component for position $j$, for $j=0, 1, \cdots, d_{emb}/2$ and $n=1,2,\cdots, N_{t_{u}}$.

\textbf{Timestep tokenization}:
Concurrently, the initial step $\hat{t}_{u}$ and the objective timestep $\hat{t}^{\text{align}}$ are both projected into timestep tokens $\bm t_{0,u}^{\text{emb}} ,~\bm t_{1,u}^{\text{emb}} \in \mathbb{R}^{1\times d_{\text{emb}}}$, to control the degree to which the diffusion models denoise. In specific, $\bm t^{\text{emb}}_{i, u}=\text{FFN}(\text{SiLU}(\text{FFN}(\bm \psi_i)))$,
where $i\in\{0,1\}$, $\text{SiLU}(\cdot)$ denotes the Sigmoid linear unit (SiLU) activation function, and $\bm \psi=\left[\sin(t\bm \omega), \cos(t\bm \omega) \right]$, with $\bm \omega[j]=\text{exp}\left(-\frac{j\cdot \log(10000)}{d_{\text{emb}}/2-1} \right)=10000^{-j/(d_{\text{emb}}/2-1)}$.
This sinusoidal encoding strategy provides a smooth and continuous representation across all timesteps. 

%The timestep $t$ is first multiplied by a frequency vector $\bm \omega \in \mathbb{C}^{d_{\text{emb}}/2}$ with
%\vspace{-0.1cm}
%\begin{equation}
%    \bm \omega[j]=\text{exp}\left(-\frac{j\cdot \log(10000)}{d_{\text{emb}}/2-1} \right)=10000^{-j/(d_{\text{emb}}/2-1)}, 
%    \vspace{-0.1cm}
%\end{equation}
%for $j=0,1,\cdots, \frac{d_{\text{emb}}}{2}-1$, followed by the concatenation $\bm \psi=\left[\sin(t\bm \omega), \cos(t\bm \omega) \right]\in \mathbb{R}^{1\times d_{\text{emb}}}$.
%Then a multi-layer perception (MLP) is used to conduct a non-linear %transformation as
%\vspace{-0.1cm}
%\begin{equation}
%    \bm t^{\text{emb}}_{i, u}=\text{FFN}(\text{SiLU}(\text{FFN}(\bm \psi))),
%    \vspace{-0.1cm}
%\end{equation}
%where $i\in\{0,1\}$ and $\text{SiLU}(\cdot)$ denotes the Sigmoid linear unit (SiLU) activation function.

%The continuous nature of this time embedding scheme facilitates stable training and smooth transitions between different noise levels, which is particularly important for wireless signal detection tasks where precise control over the denoising trajectory is essential for maintaining signal integrity and detection accuracy.

\textbf{Denoising Transformer Layer}:
One layer of Transformer encoder serves as the backbone to capture the inter-token relationship.
The position-embeded signal tokens and the embeded timesteps are concatenated and then fed into this layer, and the noise from $\hat{t}^{\text{align}}$ to $\hat{t}_{u}$ is predicted and eliminated.
\vspace{-0.1cm}
\begin{equation}
    \tilde{\bm x}_{u}^{\text{denoise}}=\!\tilde{\bm x}_{u}^{\text{PE}}\! - \!\text{Transformer}\!\left(\!\left[ \tilde{\bm x}_{u}^{\text{PE, T}}\!, \bm t_{1}^{\text{emb, T}}, \bm t_{2}^{\text{emb, T}} \right]^{\text{T}}\right).
\end{equation}

\textbf{Signal decoding}:
The denoised signals in the latent space are then projected back to the original plane as $\tilde{\bm x}_{u}^{\text{align}}=\text{FFN}(\tilde{\bm x}_{u}^{\text{denoise}})\in \mathbb{R}^{N_{t_{u}}\times 2}$.

%By measuring the equivalent SNR level and conduct the timestep-aware denoising for each user, the MU-Aligner harmonizes effective SNRs across all the users to a similar level, overcoming the multi-user imbalance and providing the foundation for the further estimation refinement.

The MU-Aligner module harmonizes the SNRs across all users to a similar level by measuring the equivalent noise power of each user and performing timestep-aware denoising accordingly. This process mitigates the multi-user imbalance issue, thereby establishing a robust foundation for subsequent estimation refinement.

\vspace{-0.2cm}
\subsection{Wireless Diffusion Transformer}
\begin{figure*}[t]
  \vspace{-0.2cm}
  \centering
  \includegraphics[width=0.90\linewidth]{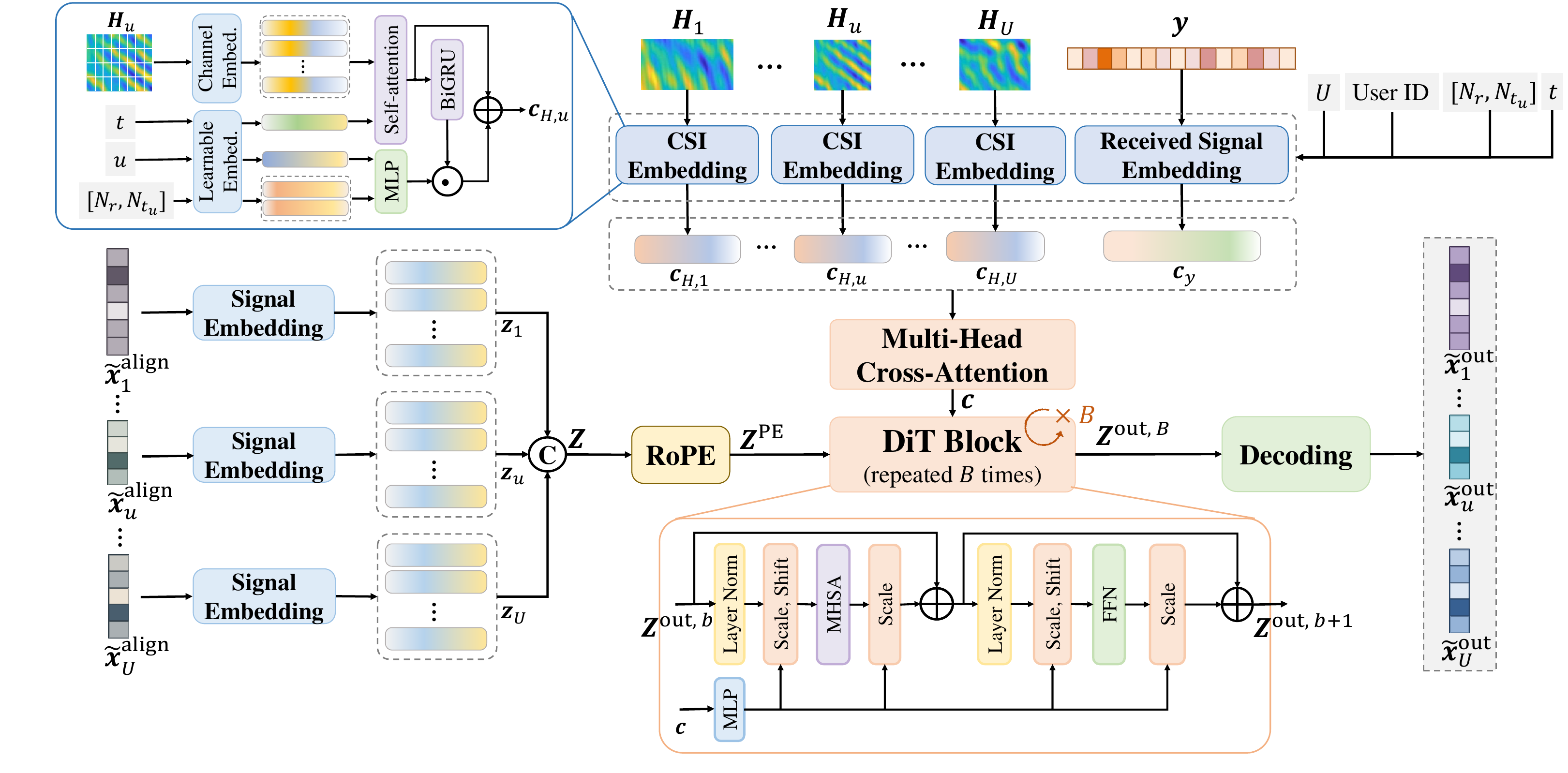}
  \vspace{-0.15cm}
  \captionsetup{font=small}
  \caption{The network architecture of WirelessDiT in WiFo-MUD.}
  \label{fig:wireless_dit}
  \vspace{-0.4cm}
\end{figure*}

The core of WiFo-MUD is the \textbf{wireless diffusion Transformer (WirelessDiT)}, a conditional denoising architecture based on the Diffusion Transformer (DiT), tailored for wireless signal demodulation in heterogeneous multi-user systems. It consists of four components: Signal embedding, condition fusion, denoising blocks and signal decoding. The aligned multi-user estimates $\{\tilde{\bm x}^{\text{align}}_{u}\}_{u=1}^{U}$ are fed into WirelessDiT and the output is the refinement of the symbols $\{\hat{\bm x}^{\text{out}}_{u}\}_{u=1}^{U}$. The overall architecture of WirelessDiT is illustrated in Fig.~\ref{fig:wireless_dit}

\subsubsection{Signal Embedding}
This module encodes the aligned estimates $\tilde{\bm{x}}_{u}^{\text{align}}$ into a high-dimensional latent representation as $\bm{z}_{u}=\text{MLP}(\tilde{\bm{x}}_{u}^{\text{align}})\in \mathbb{R}^{N_{t_{u}}\times d_{\text{emb}}}$ through a multi-layer perception (MLP).
Then the latent representations of all users are concatenated as $\bm Z= \left[\bm z_{1}^{\text{T}}, \cdots, \bm z_{U}^{\text{T}}\right]^{\text{T}}\in \mathbb{R}^{L\times d_{\text{emb}}}$, with $L=\sum_{u=1}^{U}N_{t{u}}$, followed by RoPE for the embedding of relative positional information as $\bm Z^{\text{PE}}=\text{RoPE}(\bm Z)$.

\subsubsection{Condition Fusion Module}
The latent representations are denoised guided by the conditions, including the received signal $\bm y$ and channels $\bm H_{u}$. In the condition fusion module, $\{\hat{\bm H}_{u}\}_{u=1}^{U}$ and $\bm y$ are first encoded respectively and then fused to construct the condition vector.

\textbf{CSI Embedding Module}:
We first construct a CSI embedding (CE) module $\mathcal{E}_{H}$ to pre-process the channel information, yielding the channel condition vector
\vspace{-0.2cm}
\begin{equation}
\bm c_{H,u}=\mathcal{E}_{H}\left (\hat{\bm H}_{u}, t, u, [N_{r}, N_{t_{u}}]\right ),
\vspace{-0.1cm}
\end{equation}
where the timestep $t$, user identity $u$ and the channel shape $[N_r, N_{t_{u}}]$ are embedded into the channel-related condition as side information, to help the model adapt to varying configurations. Specifically, $\hat{\bm H}_{u}$ is first divided into non-overlapping patches of size $p_{x} \times p_{y}$. These patches are embedded as $L_{H,u}=\frac{N_{r}N_{t_{u}}}{p_{x}p_{y}}$ tokens, followed by RoPE as $\bm H_{u}^{\text{emb}}=\text{RoPE}\left[ \text{MLP}\left( \text{Conv2D}(\hat{\bm H_{u}}) \right )\right ]  \in\mathbb{R}^{L_{H,u}\times d_{\text{emb}}}$.
%embedding are designed in this CSI embedding module, which can be expressed as
%\vspace{-0.1cm}
%\begin{align}
%    ~~\bm u^{\text{emb}}&=\text{MLP}(\text{Embedding(u)}),  \\
%    ~~\bm s^{\text{emb}}&=\text{MLP}\left([  \text{Embedding}(N_{t_{u}}), ~\text{Embedding}(N_{r})]\right), \\
%    ~~\bm t^{\text{emb}}&=\text{TE}(t)
%\end{align}
%where $\text{Embedding}(i)$ denotes the embedding operation for the discrete number $i$ into $d_{\text{emb}}$-dimensional space, with learnable parameters.
Concurrently, learnable embeddings for $u$, $[N_r, N_{t_{u}}]$ and $t$ are generated as $\bm u^{\text{emb}}\in \mathbb{R}^{d_{\text{emb}}}$, $\bm s^{\text{emb}}\in \mathbb{R}^{2d_{\text{emb}}}$ and $ \bm t^{\text{emb}}\in \mathbb{R}^{d_{\text{emb}}}$ respectively.
These auxiliary embeddings are combined with the channel tokens through a gated fusion mechanism. The complete CSI embedding for the $u$-th user is obtained as
\begin{equation}
\begin{aligned}
\bm c_{H,u} &=  \text{Average}\left( \text{Self-Attention}\left( [\bm H_{u}^{\text{emb}, \text{T}}, \bm t^{\text{emb}, \text{T}}]^{\text{T}} \right) \right) \\
& + \!\text{MLP}\!\left([\bm u^{\text{emb}},\bm s^{\text{emb}}] \right)\! \odot \!\text{Average}\!\left( \text{BiGRU}(\bm H_{u}^{\text{emb}}) \right),
\end{aligned}
\end{equation}
where $\text{Average}(\cdot)$ performs token-wise averaging, $\text{BiGRU}(\cdot)$ captures sequential dependencies among tokens \cite{BiGRU}. %A residual connection is employed to stabilize the aggregation process.

\textbf{Received Signal Embedding Module}:
The received $\bm y$ is pre-processed to obtain $\bm c_{y}=\mathcal{E}_{y}(\bm y, t, U, N_{r})$.
In this module, the system configuration parameters, including $U$, $N_{r}$ and $t$ are injected into signal tokens from $\bm y$ by dedicated discrete embedding modules, which operate in a similar fashion to the CSI embedding described previously. Their detailed design is therefore omitted for brevity.

\textbf{Cross-Attention Module}:
To integrate the channel-related and signal-related conditions, we fuse their feature representations via a multi-head cross-attention module as
%To extract the correlation between the channel-related and signal-related conditions, we employ a multi-head cross-attention module to fuse their features as
\vspace{-0.2cm}
\begin{equation}
\bm c=\text{CrossAttention}(\{\bm c_{H,u}\}_{u=1}^{U}, \bm c_{y})\in \mathbb{R}^{1\times d_{\text{emb}}}.
\vspace{-0.1cm}
\end{equation}
This module not only models the dependencies between the two condition types, but also captures correlations across multi-user channels, contributing to the alleviation of IUI. The resulting unified condition vector $\bm c$ is then used to guide the estimation refinement process.
%The cross-attention module can also capture the correlation between the multi-user channels, and output the final condition to guide estimation noise prediction. 

\subsubsection{Denoising Block}
This component performs the reverse diffusion process in the latent space to reconstruct the clean signal based on the coarse estimates.
At each diffusion step $t$, the output is updated through a series of WirelessDiT blocks.

First, the condition vector $\bm c$ is projected to generate modulation parameters in each WirelessDiT block:
\vspace{-0.1cm}
\begin{equation}
    \left[ \bm \gamma_{1}^{b},\bm \beta_{1}^{b}, \bm \gamma_{2}^{b}, \bm \beta_{2}^{b}, \alpha^{b} \right ]=\text{MLP}(\bm c),
    \vspace{-0.1cm}
\end{equation}
where $\bm \gamma_1^{b}, \bm \beta_1^{b},\bm \gamma_2^{b},\bm \beta_2^{b} \in \mathbb{R}^{d_{\text{emb}}}$ are scale and shift parameters in the $b$-th WirelessDiT block, and $\alpha^{b} \in \mathbb{R}$ is a gating parameter, as operated in AdaLN-Zero \cite{DiT}.
Then the input undergoes adaptive layer normalization using the condition-derived parameters as
\vspace{-0.2cm}
\begin{equation}
    \bm Z^{in,b}=\bm \gamma_{1}^{b}\odot\frac{\text{LN}(\bm Z^{\text{PE}})-\mu(\text{LN}(\bm Z^{\text{PE}}))}{\sigma(\text{LN}(\bm Z^{\text{PE}}))+\epsilon}+\bm \beta_{1}^{b},
    \vspace{-0.1cm}
\end{equation}
where $\text{LN}(\cdot)$ performs layer normalization, $\mu(\cdot)$ and $\sigma(\cdot)$ compute the mean and standard deviation along the feature dimension, $\epsilon$ is a small constant for numerical stability.
The next is a multi-head self-attention expressed as
\vspace{-0.2cm}
\begin{equation}
    \bm Z^{\text{attn},b}=\text{Softmax}\left(\frac{\bm Z^{\text{in},b}\bm W_{Q}(\bm Z^{\text{in},b}\bm W_{K})^{\text{T}}}{\sqrt{d_{qk}}}\right)\bm Z^{\text{in},b}\bm W_{v},
    \vspace{-0.1cm}
\end{equation}
where $\mathbf{W}_Q^{b}, \mathbf{W}_K^{b}, \mathbf{W}_V^{b} \in \mathbb{R}^{d_{\text{emb}} \times d_{qk}}$ are query, key, and value projection matrices in the $b$-th WirelessDiT block, and $d_{qk}$ is the dimension per attention head.
The attention output is gated using the condition-derived parameter as $\bm Z^{\text{gate},b}= \alpha^{b} \bm Z^{\text{attn},b}$, with $\alpha^{b}$ incorporating attention mechanisms.
The original input is then combined with the processed features through residual connection as $\bm Z^{\text{res}, b} = \bm Z^{\text{gate},b} + \bm Z^{\text{in},b}$.
After that, these features undergoes a second adaptive normalization and feed-forward processing as $\bm Z^{\text{ffn}, b}$.
The final output in the $b$-th WirelessDiT block is computed as $\bm Z^{\text{out}, b} = \bm Z^{\text{res}, b} + \bm Z^{\text{ffn}, b}$. 
These blocks embody the principle of gradual error correction dictated by the SDE. As the process evolves along the trajectory, the coarse estimation is successively refined, leading to a reduction in overall estimation error.

\subsubsection{Signal Decoding}
Finally, the refined latent signal is projected back to the symbol space to produce the final estimate as $\tilde{\bm x}^{\text{out}}=\text{MLP}(\bm Z^{out, B})\in \mathbb{R}^{L\times 2}$,
and the complex-valued symbols are recovers as $\hat{\bm x}^{out}= \tilde{\bm x}^{\text{out}}[:,1]+j\tilde{\bm x}^{\text{out}}[:,2]$. Based on these refined estimates, the transmitted bits are recovered via symbol quantization.

\vspace{-0.2cm}
\subsection{Model Training}

In WiFo-MUD, the MU-Aligner and WirelessDiT are pre-trained sequentially. The MU-Aligner is first trained to align the estimates across different users. Using the output of MU-Aligner as part of the input, WirelessDiT is pre-trained to further denoise the estimates.
For both modules, we adopt a self-supervised diffusion-reconstruction pre-training strategy to learn robust signal representations that are resilient to various channel impairments and interference patterns. The training process follows the DDPM framework, which consists of a forward diffusion process and a reverse denoising process.

Let $\bm x^{(0)}$ be the input symbol vector, and the conditions be $\bm c$ for simplicity.
The forward diffusion process gradually adds Gaussian noise to $\bm x^{(0)}$ over a series of $T$ timesteps $t = 1, 2, \dots, T$, producing a sequence of noisy latent variables $\bm{x}^{(1)}, \bm{x}^{(2)}, \dots, \bm{x}^{(T)}$. The transition from $\bm{x}^{(t-1)}$ to $\bm{x}^{(t)}$ is defined as
\vspace{-0.2cm}
\begin{equation}
q(\bm{x}^{(t)} | \bm{x}^{(t-1)}, \bm c) = \mathcal{N}(\bm{x}^{(t)}; \sqrt{1 - \beta_t} \bm{x}^{(t-1)}, \beta_{t} \mathbf{I}),
\vspace{-0.1cm}
\end{equation}
where $\{\beta_{t} \in (0,1)\}_{t=1}^{T}$ is a fixed variance schedule controlling the amount of noise added at each step. A notable property of this Markov chain is that any intermediate state $\bm{x}^{(t)}$ can be expressed in closed form given $\bm{x}^{(0)}$:
\vspace{-0.2cm}
\begin{equation}
\bm{x}^{(t)} = \sqrt{\bar{\alpha}_t} \bm{x}^{(0)} + \sqrt{1 - \bar{\alpha}_t} \bm{\epsilon},
\vspace{-0.1cm}
\end{equation}
where $\alpha_t = 1 - \beta_t$, $\bar{\alpha}_t = \prod_{s=1}^{t} \alpha_s$, and $\bm{\epsilon} \sim \mathcal{N}(\mathbf{0}, \mathbf{I})$ is the normal Gaussian noise.

The model, parameterized by $\theta$ for simplicity, is trained to reverse this diffusion process by predicting the noise vector $\bm{\epsilon}$ from the noisy input $\bm {x}^{(t)}$. 
%The training is conditioned on the channel matrix $\mathbf{H}$, the received signal $\mathbf{y}$, and the user ID to incorporate communication-specific context. 
The training objective is to minimize the mean-squared error between the predicted and true noise, i.e., to optimize
\vspace{-0.2cm}
\begin{equation}
\mathcal{L}(\theta) = \mathbb{E}_{t, \bm{x}^{(0)}, \bm{\epsilon}} \left[ \| \bm{\epsilon} - \bm{\epsilon}_\theta( \mathbf{x}^{(t)}|t , \bm c) \|_{2}^2 \right],
\vspace{-0.2cm}
\end{equation}
where $t$ is uniformly sampled from $\{1, 2, \dots, T\}$, $\bm{x}^{(0)}$ is drawn from the training dataset, and $\bm{\epsilon} \sim \mathcal{N}(\mathbf{0}, \mathbf{I})$. $\bm c$ includes the received signal $\bm y$ and the estimated channels $\{\hat{\bm H}_{u}\}_{u=1}^{U}$. The training algorithm is summarized in Algorithm~\ref{alg:training}.

\begin{algorithm}[h]
  \caption{Diffusion-Reconstruction Training Algorithm}
  \label{alg:training}
  \textbf{Input}: Training dataset $\mathcal{D}$, containing symbol vectors $\bm x^{(0)}$ and corresponding conditions $\bm c$; Total number of sampling timesteps $T$; Number of training epochs $E_{\text{diff}}$\\
  \textbf{Output}:Learned denoising model $\bm{\epsilon}_\theta$; \\
  \textbf{Steps}: \\
  For $n=1$ to $E_{\text{diff}}$: \\
  \transparent{0.0}{00}\transparent{1}
  Sample $t \sim \text{Uniform}({1, \dots, T})$;\\
  \transparent{0.0}{00}\transparent{1}
  Sample noise $\bm{\epsilon} \sim \mathcal{N}(\bm{0}, \bm{I})$;\\
  \transparent{0.0}{00}\transparent{1}
  Compute $\bm{x}^{(t)} = \sqrt{\bar{\alpha}_t} \bm {x}^{(0)} + \sqrt{1 - \bar{\alpha}_t} \bm{\epsilon}$;\\
  \transparent{0.0}{00}\transparent{1}
  Take gradient step on $\nabla_\theta | \bm{\epsilon} - \bm{\epsilon}_\theta( \mathbf{x}^{(t)}|  t , \bm c ) |^2$; \\
  \transparent{0.0}{00}\transparent{1}
  Update $\bm{\epsilon}_\theta$ based on the gradient.
\end{algorithm}

Note that the MU-Aligner and WirelessDiT are trained in similar manners, with different input conditions. The MU-Aligner is conditioned by both initial and objective timesteps, while the WirelessDiT is conditioned by various configuration information and a single timestep. During the training of WirelessDiT, the pre-trained MU-Aligner module remains frozen.
%To ensure robust generalization across diverse communication scenarios, the model is trained on a wide variety of modulation schemes (QPSK/8PSK/16QAM/64QAM), user counts $U$, and antenna configurations $N_{t_{u}}$ and $N_{r}$. 

\section{Inference Refinement}

Upon completion of training, WiFo-MUD generates refined bit estimates from initial coarse inputs. Its practical deployment, however, faces two primary challenges: inference latency incompatible with physical-layer frames and accuracy degradation from inter-user interference. To address these we optimize the inference in two ways. First, a communication-aware consistency distillation technique accelerates the denoising process. Second, a dynamic user-grouping scheme applies successive interference cancellation to mitigate interference.

%\textcolor{blue}{Once upon training, WiFo-MUD iteratively refines initial coarse estimates of all users to recover transmitted bits. However, practical deployment may encounter several challenges. First, the substantial inference latency of the model is incompatible with existing physical-layer frame structures, necessitating acceleration of the inference process. Effectively distilling the model while maintaining its detection accuracy remains an unresolved issue. Moreover, the accuracy of WiFo-MUD may be deteriorated by severe IUI, particularly in scenarios with dense user access, and the SNR gap across users may be too large to be aligned in some difficult scenarios. 
%}

%\textcolor{blue}{
%In this section, we propose to optimize the model inference in two directions. First, we propose a communication-aware consistency distillation to refine the model inference while eliminating the need of iterative denoising in WirelessDiT. Second, we devise a dynamic user grouping scheme in inference, leveraging successive interference cancellation to mitigate inter-user interference.}

\vspace{-0.2cm}
\subsection{Communication-Aware Consistency Distillation}

To enhance the inference speed of WiFo-MUD, we develop a \textbf{communication-aware consistency distillation} strategy. This approach transfers knowledge from the pre-trained WirelessDiT model to a more efficient student network, aiming to approximate multi-step sampling through single-step forward inference while maintaining demodulation accuracy.

%To address the high computational cost of iterative denoising during inference, we employ \textbf{consistency distillation} (CD) to transfer the knowledge from a pre-trained WirelessDiT model into an efficient student network. This approach aims to approximate the multi-step sampling process with a single forward pass, while maintaining high estimation accuracy.

We distill the pre-trained teacher WirelessDiT, $\mathcal{T}_{\bm \phi}(\tilde{\bm x}^{\text{align}, (t)}; t, \bm c)$, into a student model $\mathcal{S}_{\bm \theta}(\tilde{\bm x}^{\text{align}, (t)}; t, \bm c)$, which is designed to directly map any noisy state $\{\tilde{\bm x}^{\text{align}, (t)}\}_{u=1}^{U}$ at timestep $t$ to the corresponding clean symbols $\tilde{\bm x}^{(0)}$.
We formulate the distillation optimization using a multi-objective loss function as
\vspace{-0.1cm}
\begin{equation}
    \mathcal{L}(\bm \theta)= \eta_d  \mathcal{L}_{\text{CD}}(\bm \theta) + (1-\eta_{d}) \mathcal{L}_{\text{MSE}}(\bm \theta),
    \vspace{-0.1cm}
\end{equation}
integrating both the consistency distillation loss and the estimation MSE loss, adaptively weighted according to prevailing channel conditions.

\subsubsection{Consistency Distillation Loss}

%The key principle behind consistency distillation is to enforce that the model produces consistent outputs across points on the same probability trajectory, that is, for any two timesteps $t > s \geqslant 0$, the student model should satisfy $\mathcal{S}_{\theta}(\tilde{\bm x}^{\text{align},(t)}; t, \bm c)=\mathcal{S}_{\theta}(\tilde{\bm x}^{\text{align},(s)}; s, \bm c)$.
%To achieve this, we minimize the consistency distillation loss defined as
Similar to \cite{CM}, the consistency distillation loss is defined as
\begin{align}
\vspace{-0.2cm}
    \mathcal{L}_{\text{CD}}(\bm \theta)& =\mathbb{E}_{t} \large[ \lambda(t_{n})\|\mathcal{S}_{\theta}(\tilde{\bm x}^{\text{align},(t_{n+1})}; t_{n+1}, \bm c) \\ 
    &- \mathcal{S}_{\theta^{-}}(\tilde{\bm x}^{\text{align},(t_{n})}; t_{n}, \bm c)\|_{\text{F}}\large],  \nonumber 
\end{align}
where $\lambda(t)$ is a timestep-dependent weight that prioritizes different noise levels.
$\tilde{\bm x}^{\text{align},(t_{n})}$ is obtained by applying one denoising step to $\tilde{\bm x}^{\text{align},(t_{n+1})}$ using the teacher model. $\bm \theta^{-}$ is an exponential moving average (EMA) of student parameters $\bm \theta$ for stable training. This loss term promotes consistent model outputs for samples on the same probability trajectory.

\subsubsection{Communication-Aware Loss Design}
To further enhance data efficiency and improve demodulation stability during distillation, we incorporate domain knowledge through a refined loss function. The student model's estimated symbols are compared with the ground truth to compute the MSE loss as
\vspace{-0.1cm}
\begin{equation}
    \mathcal{L}_{\text{MSE}}(\bm \theta)=\mathbb{E} \|  \mathcal{S}_{\theta}(\!\tilde{\bm x}^{\text{align},(t_{n+1})}; t_{n+1}, \bm c) -\tilde{\bm x}^{(0)}\|_{2}^{2}.
    \vspace{-0.1cm}
\end{equation}
To optimally balance the different loss components, we design an adaptive weight based on channel conditions as
\vspace{-0.1cm}
\begin{equation}
    \eta_{d}= \frac{1}{1 + \exp\{-k_{0}( \frac{\sum_{u=1}^{U}\|\hat{\bm H}_{u}\|_{\text{F}}^{2}}{U\sigma_{n}^{2}} - s_{0} )\} },
    \vspace{-0.1cm}
\end{equation}
with $k_{0}$ and $s_{0}$ being positive constants. When channel conditions deteriorate, greater emphasis is placed on minimizing the MSE loss, while higher SNR scenarios prioritize the consistency distillation loss optimization.

The student model $\mathcal{S}_\theta$ shares the same architecture with the teacher but with separate parameters.
The distilled WirelessDiT $\mathcal{S}_\theta$ achieves comparable accuracy to the iterative teacher model while reducing inference time from $N$ steps to a single step, making it feasible for real-time systems with low latency budgets.
Algorithm~\ref{alg:cd} summarizes the complete communication-aware consistency distillation procedure.

\begin{algorithm}[h]
\caption{Communication-Aware Consistency Distillation}
\label{alg:cd}
\textbf{Input}: Dataset $\mathcal{D}$; Pre-trained teacher wirelessDiT model $\mathcal{T}_{\bm \phi}$; $\{\lambda(t_{n})\}_{n=1}^{T}$; EMA updating weight $\mu$; Number of distillation epochs $E_{\text{dis}}$  \\
\textbf{Output}: Distilled wirelessDiT model $\mathcal{S}_{\bm \theta}$ \\
%\begin{algorithmic}\\
\textbf{\!\!\!\!\!Steps:} \\
\!\!\!\!\!Initialize student parameters $\bm \theta \leftarrow \bm \theta_0$; \\
\!\!\!\!\!Initialize EMA parameters $\bm \theta^{-} \leftarrow \bm \theta$; \\
\textbf{\!\!\!\!\!Repeat} \\
\transparent{0.0}{00}\transparent{1}
Sample $\{\tilde{\bm{x}}^{\text{align}, (0)}, \mathbf{c}\}$ from $\mathcal{D}$; \\
\transparent{0.0}{00}\transparent{1}
Sample $n \sim \mathcal{U}[1, T]$;\\
\transparent{0.0}{00}\transparent{1}
Generate $\tilde{\bm{x}}^{\text{align}, (t_{n+1})}\sim \mathcal{N}(\tilde{\bm{x}}^{\text{align}, (0)}; t_{n+1}^{2}\bm I)$; \\
\transparent{0.0}{00}\transparent{1}
Compute teacher outputs: $\tilde{\bm{x}}^{\text{align}, (t_{n})}\leftarrow \tilde{\bm{x}}^{\text{align}, (t_{n+1})} - (t_{n}-t_{n+1})t_{n+1}\mathcal{T}_{\bm \phi}(\tilde{\bm{x}}^{\text{align}, (t_{n+1})}, t_{n+1}, \bm{c})$; \\
\transparent{0.0}{00}\transparent{1}
Compute the combined loss $\mathcal{L}(\bm \theta)$ as Eq.~(40); \\
\transparent{0.0}{00}\transparent{1}
Update $\bm \theta$ using Adam optimizer: $\bm \theta \leftarrow \bm \theta - \eta \nabla_{\theta} \mathcal{L}_{\text{CD}}(\bm \theta)$; \\
\transparent{0.0}{00}\transparent{1}
Update EMA: $\bm \theta^{-} \leftarrow \text{stopgrad}(\mu \bm \theta^{-} + (1-\mu) \bm \theta) $;\\ 
\textbf{\!\!\!\!\!until} convergence \textbf{or} $E_{\text{dis}}$ is reached
%\end{algorithmic}
\end{algorithm}

\vspace{-0.2cm}
\subsection{Dynamic User Grouping and Successive Inference Strategy}

To effectively mitigate the inter-user interference, we introduce a \textbf{dynamic user grouping} and successive inference cancellation strategy.

The process begins by ranking all users according to their channel strengths. The channel strength is measured by $h_{u}=\frac{\|\hat{\bm H}_{u}\|_{\text{F}}^{2}}{N_{r}N_{t_{u}}}$.
Users are sorted in descending order to form an ordered sequence as
\vspace{-0.2cm}
\begin{equation}
\Pi = \text{argsort}\left( \{ h_{1}, h_{2}, \cdots, h_{U} \}\right).
\vspace{-0.1cm}
\end{equation}
Users are divided into several groups, and users with stronger channel conditions are prioritized to be processed by WiFo-MUD, improving demodulation reliability and facilitating more accurate interference cancellation in subsequent stages.

To avoid the low-rank channel conditions, the total number of data streams should not exceed the threshold $N_{\text{thres}}$, with $N_{\text{thres}}\leqslant N_{r}$. Therefore, the group is divided recursively and selected in a greedy manner. The $i$-th group of user indices is defined as $\mathcal{G}^{(i)}$, and the user number of this group can be expressed as
\vspace{-0.2cm}
\begin{equation}
    \vert \mathcal{G}^{(i)} \vert = \!\!\!\!\!\!\underset{\sum_{l=1+\sum_{j=1}^{i-1}\vert \mathcal{G}^{(j)}\vert}^{k+\sum_{j=1}^{i-1}\vert \mathcal{G}^{(j)}\vert}  N_{t_{\Pi[l]}}  \leqslant N_{\text{thres}}}{\text{argmax}}~ \!\!\!\!  k.
    \vspace{-0.2cm}
\end{equation}
In other words, the users are chosen into the $i$-th group in descending order from the remaining set of $\Pi$, while the total number of data streams would not exceed the threshold.
The corresponding channel subset and transmitted symbol matrix for group $i$ are $\{\hat{\bm H}_{j}\}_{j\in \mathcal{G}^{(i)} }$ and $\{\hat{\bm x}_{j}\}_{j\in \mathcal{G}^{(i)} }$, respectively.

Demodulation is performed sequentially across groups. For group $i$, the received signal is calculated as
\vspace{-0.2cm}
\begin{equation}
    \bm y^{(i)} = \bm y^{(0)} - \sum_{j=1}^{i-1} \sum_{u\in{\mathcal{G}^{(j)}}}\hat{\bm H}_{u}\hat{\bm x}_{u}^{\text{out}},
    \vspace{-0.2cm}
    \label{equ:SIC}
\end{equation}
where $\bm y^{(0)}=\bm y$ and $\hat{\bm x}_{u}^{\text{out}}$ is the refined output of WiFo-MUD in former $i-1$ groups. Under the assumption that contributions from later groups act as interference, the WiFo-MUD estimates the symbols for group $i$ in one-step inference as
\begin{equation}
    \{ \hat{\bm x}_{u}^{\text{out}}\}_{u \in \mathcal{G}^{(i)}} =\text{WiFo-MUD}( \hat{\bm x}_{u}, \bm c),
\end{equation}
where $\bm c$ includes $\{\hat{\bm H}_{j}\}_{j\in \mathcal{G}^{(i)} }$, $\bm y^{(i)}$ and other configuration-related information.
After estimating the symbols and channel for group $i$, their contribution is also removed from the current received signal, similar to Eq.~(\ref{equ:SIC}), to reduce interference for later groups. This loop continues until all users are grouped and processed.

\section{Experiments}

In this section, we present a comprehensive evaluation of the proposed WiFo-MUD framework. We first describe the datasets used for model training and testing, introduce the baseline methods for comparison, and define the evaluation metrics. Then we evaluate the performance of the proposed WiFo-MUD approach from multiple perspectives.

\vspace{-0.3cm}

\subsection{Dataset Construction}

To ensure a comprehensive evaluation of the proposed model, a large-scale heterogeneous multi-user signal demodulation dataset is constructed, integrating both simulated and real-world measurements. This dataset covers a wide range of transmission scenarios, system configurations, and modulation schemes to guarantee diversity and fidelity. The channel data is generated following the methodology outlined in \cite{WiFo_CF} and consists of the following components:
\begin{itemize}
    \item \textbf{Statistical channel modeling}: This subset is generated using \textit{QuaDriGa} \cite{quadriga} following 3GPP 38.901 standards, including urban macro (UMa), urban micro (UMi), rural macro (RMa), and indoor scenarios with both LoS and NLoS conditions.
    \item \textbf{Ray-tracing}: High-fidelity channel data are incorporated from the \textit{DeepMIMO} dataset \cite{deepMIMO} and the \textit{SynthSoM} dataset \cite{SynthSoM}, produced via Wireless InSight for detailed ray-tracing simulations across diverse scenarios.
    \item \textbf{Real-world measurements}: Real-measured channel data are drawn from the \textit{Argos} dataset \cite{dataset_argos}, collected in indoor and outdoor environments across a university campus, and the \textit{Dichasus} dataset \cite{dataset_dich} covering classrooms, factories, and outdoor areas.
\end{itemize}

Our dataset comprises a balanced distribution of modulation schemes (QPSK, 8PSK, 16QAM, 64QAM), with an equal number of samples for each scenario. Each instance includes the transmitted symbols $\{\bm x_{u}\}_{u=1}^{U}$, received signal $\bm y$, channel matrix $\{\bm H_{u}\}_{u=1}^{U}$, and noise variance $\sigma_{n}^2$. The naming convention and channel-related parameter configurations of our dataset is also consistent with \cite{WiFo_CF}.
%Detailed parameter configurations are provided in Table.~\ref{tab:dataset}.

\vspace{-0.2cm}
\begin{table}[h]
\centering
\caption{Hyper-parameters for model training}
\label{tab:hyper}
\renewcommand\arraystretch{1.3}
%\begin{tabular}{|cc|cc|cc}
\begin{tabular}{@{} p{0.18cm} p{0.76cm} |p{0.18cm} p{0.76cm}| p{0.18cm} p{0.65cm}@{}}
\toprule
\multicolumn{2}{c|}{MU-Aligner}       & \multicolumn{2}{c|}{WirelessDiT}   & \multicolumn{2}{c}{Model Distillation}   \\  
\midrule  \midrule
\multicolumn{1}{c}{Epochs}  & 100 & \multicolumn{1}{c}{Epochs}  &  100 & \multicolumn{1}{c}{Epochs}   & 80  \\ \hline
\multicolumn{1}{c}{Batch size}    & 64  & \multicolumn{1}{c}{Batch size}  &64  & \multicolumn{1}{c}{Batch size}  & 64   \\ \hline
\multicolumn{1}{c}{$L_r$} & 0.002  & \multicolumn{1}{c}{$L_r$} &0.001 & \multicolumn{1}{c}{$L_r$}  & 0.0008 \\ \hline
\multicolumn{1}{c}{Optimizer}   & $\text{AdamW}$ & \multicolumn{1}{c}{Optimizer}   &  $\text{AdamW}$ & \multicolumn{1}{c}{Optimizer}   &  $\text{Adam}$  \\ \hline
\multicolumn{1}{c}{Weight decay}   & 0.0001  & \multicolumn{1}{c}{Timesteps}   & 1000 & \multicolumn{1}{c}{Weight $\mu$}   &  0.2  \\
\bottomrule
\end{tabular}
\end{table}
\vspace{-0.2cm}

\subsection{Comparison Methods}

\begin{table*}[b]
\centering
\caption{Network architecture parameters of WiFo-MUD with varying sizes.}
\label{tab:three}
\renewcommand\arraystretch{1.3}
\begin{tabular}{cccccccc}
\toprule
Models         & Aligner Depth & Aligner Width & Aligner Heads & WirelessDiT   Depth & WirelessDiT   Width & WirelessDiT   Heads & Parameters \\ 
\midrule
WiFo-MUD-Small & 1             & 64            & 4             & 3                   & 64                  & 8                   & 1.29M      \\
WiFo-MUD-Base  & 1             & 64            & 4             & 6                   & 128                 & 8                   & 3.92M      \\
WiFo-MUD-Large & 2             & 64            & 4             & 8                   & 256                 & 8               & 5.83M      \\ 
\bottomrule
\end{tabular}
\end{table*}

\begin{figure}[t]
  \vspace{-0.1cm}
  \centering
  \includegraphics[width=0.9\linewidth]{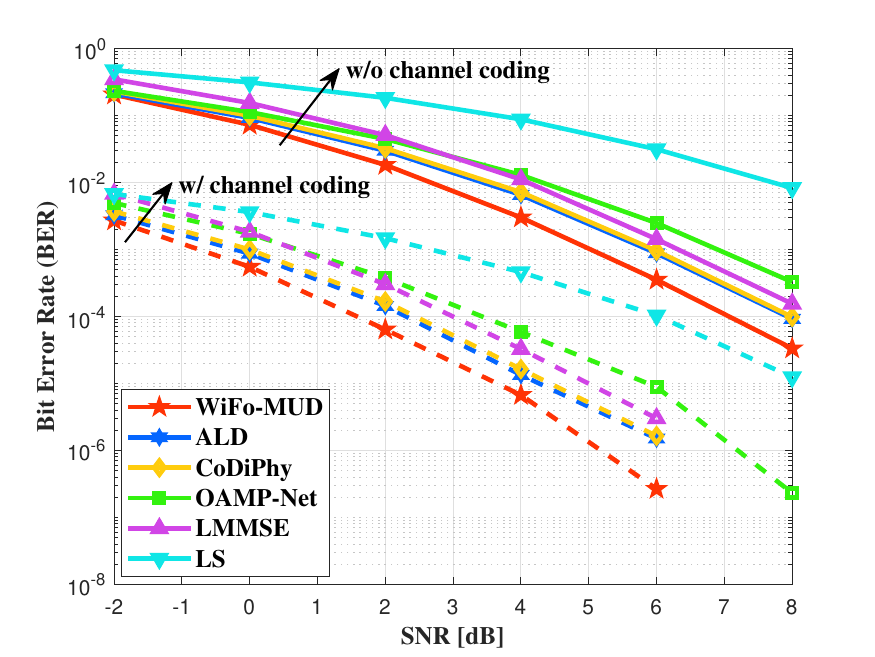}
  \vspace{-0.1cm}
  \captionsetup{font=small}
  \caption{BER comparison under varying SNR among different demodulators.}
  \label{fig:ber}
  \vspace{-0.45cm}
\end{figure}

\begin{figure}[t]
  \vspace{-0.1cm}
  \centering
  \includegraphics[width=0.9\linewidth]{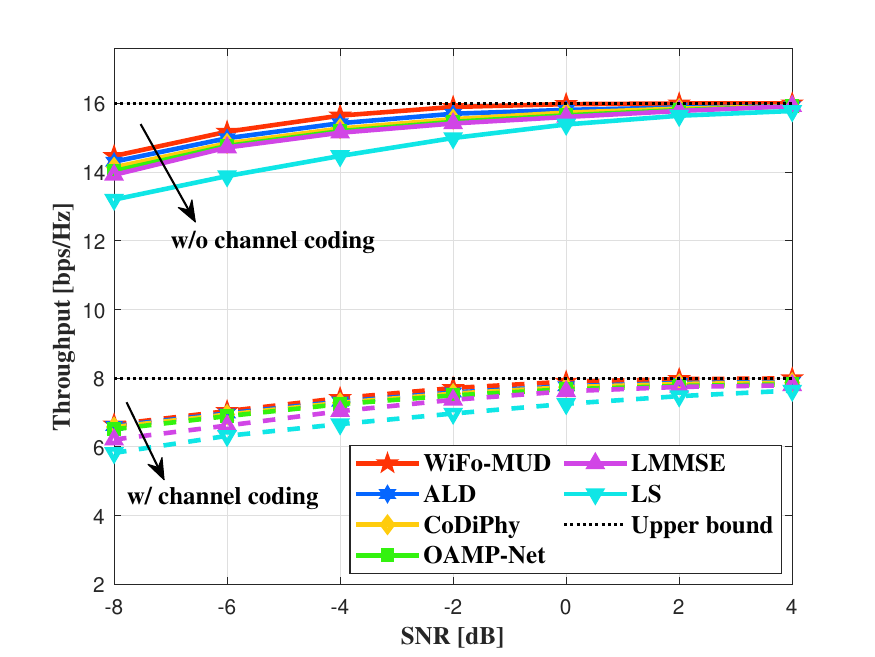}
  \vspace{-0.1cm}
  \captionsetup{font=small}
  \caption{Throughput comparison under varying SNR among different demodulators.}
  \label{fig:throughput}
  \vspace{-0.3cm}
\end{figure}

In our experiments, we developed three architectural variants of the proposed WiFo-MUD model, termed \textbf{WiFo-MUD-Large}, \textbf{WiFo-MUD-Base}, and \textbf{WiFo-MUD-Small}, which vary in the number of network layers and the dimensionality of the latent space. The training hyper-parameters adopted for the WiFo-MUD models are provided in Table~\ref{tab:hyper}, while detailed architectural specifications are summarized in Table~\ref{tab:three}.
We compare WiFo-MUD against the following representative baseline methods:

\begin{itemize}
\item \textit{Linear Demodulators}: LS and LMMSE estimators.
\item \textit{OAMP-Net}~\cite{OAMPNet}: A deep learning demodulator that unrolls the orthogonal approximate message passing algorithm into a trainable network, with learnable damping and divergence parameters optimized through end-to-end training.
\item \textit{ALD}~\cite{ALD_2023TWC}: A diffusion-based demodulator employing annealed Langevin dynamics, where symbols are estimated via annealing sampling in the posterior score using pre-trained channel-specific score networks.
\item \textit{CoDiPhy}~\cite{CoDiPhy}: A conditional DDPM-based framework for signal demodulation that requires independent training for each system configuration due to limited adaptability to heterogeneous scenarios.
\end{itemize}
All learning-based benchmark models are trained using their officially recommended configurations.

% Please add the following required packages to your document preamble:
% \usepackage{multirow}
% \usepackage[table,xcdraw]{xcolor}
% Beamer presentation requires \usepackage{colortbl} instead of \usepackage[table,xcdraw]{xcolor}
\begin{table*}[t]
\renewcommand\arraystretch{1.3}
\caption{\textbf{Full-shot} BER comparison among different demodulators. The \colorbox[HTML]{FFCCCB}{Light red} shading indicates the highest accuracy, and \colorbox[HTML]{D4F1F9}{light blue} highlights the next best performance.}
\label{tab:full_shot}
\begin{tabular}{ccccccccccccc}
\toprule
& \multicolumn{2}{c}{\textbf{Ours}} & \multicolumn{2}{c}{\textbf{ALD}}  & \multicolumn{2}{c}{\textbf{CoDiPhy}} &\multicolumn{2}{c}{\textbf{OAMP-Net}} & \multicolumn{2}{c}{\textbf{LMMSE}}   & \multicolumn{2}{c}{\textbf{LS}}  \\ \cline{2-13} 
\multirow{-2}{*}{\textbf{Dataset}} & \multicolumn{1}{l}{w/o cod.}   & w/ cod.  & \multicolumn{1}{l}{w/o cod.}   & w/ cod.   & \multicolumn{1}{l}{w/o cod.}   & w/ cod.   & \multicolumn{1}{l}{w/o cod.}   & \multicolumn{1}{l}{w/ cod.} & \multicolumn{1}{l}{w/o cod.}   & \multicolumn{1}{l}{w/ cod.} & \multicolumn{1}{l}{w/o cod.}   & w/ cod.   \\ 
\midrule  \midrule
\textbf{Q1}  & \cellcolor[HTML]{FFCCCB}\textbf{0.0782}  &\cellcolor[HTML]{FFCCCB}\textbf{0.0005}   & \cellcolor[HTML]{D4F1F9}0.0803 & \cellcolor[HTML]{D4F1F9}0.0006   & 0.0835 & 0.0007  & 0.0832 &  0.0007   & 0.1212 &  0.0013 & 0.1366 &  0.0014 \\

\textbf{Q2}  & \cellcolor[HTML]{FFCCCB}\textbf{0.0809} &\cellcolor[HTML]{FFCCCB}\textbf{0.0005}   & \cellcolor[HTML]{D4F1F9}0.0832 & \cellcolor[HTML]{D4F1F9}0.0006  & 0.0856 &  0.0008  & 0.0872  &  0.0009    & 0.1202 &  0.0012 & 0.1347 &   0.0013    \\

\textbf{Q3}  & \cellcolor[HTML]{FFCCCB}\textbf{0.0898} &\cellcolor[HTML]{FFCCCB}\textbf{0.0006} & \cellcolor[HTML]{D4F1F9}0.0913  &  \cellcolor[HTML]{D4F1F9}0.0007   & 0.0941     &  0.0009 & 0.0940  & 0.0009   & 0.1230 &  0.0013  & 0.1387  &  0.0015  \\

\textbf{Q4}  & \cellcolor[HTML]{FFCCCB}\textbf{0.0781}  & \cellcolor[HTML]{FFCCCB}\textbf{0.0004} & \cellcolor[HTML]{D4F1F9}0.0798 &  \cellcolor[HTML]{D4F1F9}0.0006  & 0.0803      & 0.0007 & 0.0832   &  0.0008  & 0.1092  &  0.0011  & 0.1181  & 0.0013   \\

\textbf{Q5}  & \cellcolor[HTML]{FFCCCB}\textbf{0.0790} & \cellcolor[HTML]{FFCCCB}\textbf{0.0005}   & \cellcolor[HTML]{D4F1F9}0.0812   &  \cellcolor[HTML]{D4F1F9}0.0007    & 0.0819   & \cellcolor[HTML]{D4F1F9}0.0007  & 0.0831  & 0.0008   & 0.1216  & 0.0012 & 0.1306 & 0.0013  \\

\textbf{Q6} & \cellcolor[HTML]{FFCCCB}\textbf{0.0782}   &  \cellcolor[HTML]{FFCCCB}\textbf{0.0005}  &\cellcolor[HTML]{D4F1F9}~~0.0801   &  \cellcolor[HTML]{FFCCCB}\textbf{0.0005} & 0.0818 &  \cellcolor[HTML]{D4F1F9}0.0007   & 0.0822     &  0.0008    & 0.1210 &  0.0012 & 0.1287 & 0.0014   \\

\textbf{Q7}  & \cellcolor[HTML]{FFCCCB}\textbf{0.0773}  & \cellcolor[HTML]{FFCCCB}\textbf{0.0004}  & \cellcolor[HTML]{D4F1F9}0.0811  & \cellcolor[HTML]{D4F1F9}0.0006  & 0.0835  &  0.0008    & 0.0840   &  0.0009   & 0.1241 &  0.0013 & 0.1311  & 0.0014  \\

\textbf{Q8} & \cellcolor[HTML]{FFCCCB}\textbf{0.0819}    & \cellcolor[HTML]{FFCCCB}\textbf{0.0005}  &0.0860  & 0.0007  & \cellcolor[HTML]{D4F1F9}0.0854  & \cellcolor[HTML]{D4F1F9}0.0006   & 0.0892   &   0.0007  & 0.1160  &  0.0011 & 0.1252   & 0.0014 \\

\textbf{D1} & \cellcolor[HTML]{FFCCCB}\textbf{0.0881}  & \cellcolor[HTML]{FFCCCB}\textbf{0.0006}  & \cellcolor[HTML]{D4F1F9}0.0898    &  \cellcolor[HTML]{D4F1F9}0.0008  & 0.0908  & \cellcolor[HTML]{D4F1F9}0.0008  & 0.0952 &  0.0009   & 0.1186  &  0.0013  & 0.1365  & 0.0015 \\

\textbf{D2} & \cellcolor[HTML]{FFCCCB}\textbf{0.0870}   & \cellcolor[HTML]{FFCCCB}\textbf{0.0006}  & 0.0901   & 0.0008   & \cellcolor[HTML]{D4F1F9}0.0896  &  \cellcolor[HTML]{D4F1F9}0.0007   & 0.0975  &  0.0009  & 0.1196 & 0.0012  & 0.1329  &  0.0014 \\ \midrule
\textbf{Ave.}  & \cellcolor[HTML]{FFCCCB}\textbf{0.0819} & \cellcolor[HTML]{FFCCCB}\textbf{0.0005} & \cellcolor[HTML]{D4F1F9}0.0843 & \cellcolor[HTML]{D4F1F9}0.0007 & 0.0857 &  \cellcolor[HTML]{D4F1F9}0.0007 & 0.0879 &  0.0008  & 0.1195 &  0.0012   & 0.1313 & 0.0014 \\ \bottomrule
\end{tabular}
\end{table*}

% Please add the following required packages to your document preamble:
% \usepackage{multirow}
% \usepackage[table,xcdraw]{xcolor}
% Beamer presentation requires \usepackage{colortbl} instead of \usepackage[table,xcdraw]{xcolor}
\begin{table*}[]
\caption{\textbf{Zero-shot} BER comparison among different demodulators. The \colorbox[HTML]{FFCCCB}{Light red} shading indicates the highest accuracy, and \colorbox[HTML]{D4F1F9}{light blue} highlights the next best performance.}
\label{tab:zero_shot}
\renewcommand\arraystretch{1.3}
\begin{tabular}{ccccccccccccc}
\toprule
& \multicolumn{2}{c}{\textbf{Ours}}   & \multicolumn{2}{c}{\textbf{ALD}} & \multicolumn{2}{c}{\textbf{CoDiPhy}} &\multicolumn{2}{c}{\textbf{OAMP-Net}}  & \multicolumn{2}{c}{\textbf{LMMSE}} & \multicolumn{2}{c}{\textbf{LS}} \\ \cline{2-13} 
\multirow{-2}{*}{\textbf{Dataset}} & \multicolumn{1}{l}{w/o cod.}   & w/ cod.   & \multicolumn{1}{l}{w/o cod.}   & w/ cod.  & \multicolumn{1}{l}{w/o cod.}   & w/ cod. & \multicolumn{1}{l}{w/o cod.}   & \multicolumn{1}{l}{w/ cod.} & \multicolumn{1}{l}{w/o cod.}   & \multicolumn{1}{l}{w/ cod.} & \multicolumn{1}{l}{w/o cod.}   & w/ cod.  \\ 
\midrule \midrule
\textbf{Q9} & \cellcolor[HTML]{FFCCCB}\textbf{0.0891} & \cellcolor[HTML]{FFCCCB}\textbf{0.0007}   & \cellcolor[HTML]{D4F1F9}0.0907 &  \cellcolor[HTML]{D4F1F9}0.0008  & 0.0923 &  \cellcolor[HTML]{D4F1F9}0.0008    & 0.1542&  0.0016   & 0.1218  &  0.0013  & 0.1320 & 0.0015    \\

\textbf{Q10} & \cellcolor[HTML]{FFCCCB}\textbf{0.0896} &  \cellcolor[HTML]{FFCCCB}\textbf{0.0007}  & \cellcolor[HTML]{D4F1F9}0.0899 &   \cellcolor[HTML]{D4F1F9}0.0008  & 0.0908 &  \cellcolor[HTML]{D4F1F9}0.0008   & 0.1239 & 0.0014    & 0.1227 &  0.0014  & 0.1357 &  0.0015    \\

\textbf{Q11} & \cellcolor[HTML]{FFCCCB}\textbf{0.0903} & \cellcolor[HTML]{FFCCCB}\textbf{0.0008}  & \cellcolor[HTML]{D4F1F9}0.0903  &   \cellcolor[HTML]{D4F1F9}0.0009  & 0.0907  &  \cellcolor[HTML]{D4F1F9}0.0009   & 0.1315  & 0.0016 & 0.1272 &  0.0015  & 0.1408 &   0.0016  \\

\textbf{Q12} & \cellcolor[HTML]{FFCCCB}\textbf{0.0910} &  \cellcolor[HTML]{FFCCCB}\textbf{0.0008}  & \cellcolor[HTML]{D4F1F9}0.0914 &   \cellcolor[HTML]{D4F1F9}0.0009  & 0.0955  & 0.0010  & 0.1402  &  0.0015   & 0.1197 &  0.0013  & 0.1337 &  0.0015     \\

\textbf{Q13}& \cellcolor[HTML]{FFCCCB}\textbf{0.0862}   & \cellcolor[HTML]{FFCCCB}\textbf{0.0008}   & \cellcolor[HTML]{D4F1F9}0.0877 &  \cellcolor[HTML]{D4F1F9}0.0010  & 0.0895  &  \cellcolor[HTML]{D4F1F9}0.0010  & 0.1104  & 0.0012    & 0.1178 &   0.0013 & 0.1289 &   0.0015    \\

\textbf{Q14} & \cellcolor[HTML]{FFCCCB}\textbf{0.0859} &  \cellcolor[HTML]{FFCCCB}\textbf{0.0007} & \cellcolor[HTML]{D4F1F9}0.0861 &  \cellcolor[HTML]{FFCCCB}\textbf{0.0007} & 0.0865   &  \cellcolor[HTML]{D4F1F9}0.0009  & 0.0998  & 0.0011   & 0.1089 &   0.0012 & 0.1295  &  0.0014    \\

\textbf{Q15}  & \cellcolor[HTML]{D4F1F9}0.0924 &  \cellcolor[HTML]{FFCCCB}\textbf{0.0009} & \cellcolor[HTML]{FFCCCB}\textbf{0.0921} & \cellcolor[HTML]{FFCCCB}\textbf{0.0009}    & 0.0984  &  \cellcolor[HTML]{D4F1F9}0.0010 & 0.1087   &   0.0011 & 0.1106  &  0.0013  & 0.1298 &    0.0014   \\

\textbf{Q16}  & \cellcolor[HTML]{FFCCCB}\textbf{0.0906} &  \cellcolor[HTML]{FFCCCB}\textbf{0.0008}  & \cellcolor[HTML]{D4F1F9}0.0909  &  \cellcolor[HTML]{D4F1F9}0.0009  & 0.0918  & 0.0010  & 0.1154  &   0.0011  & 0.1160  &  0.0012 & 0.1307 &   0.0015   \\

\textbf{D3}  & \cellcolor[HTML]{FFCCCB}\textbf{0.0934} & \cellcolor[HTML]{FFCCCB}\textbf{0.0009}  & \cellcolor[HTML]{D4F1F9}0.0946 &  \cellcolor[HTML]{D4F1F9}0.0011   & 0.0968  & 0.0012  & 0.1187  & 0.0014     & 0.1204 & 0.0014  & 0.1358 &  0.0016    \\

\textbf{D4}  & \cellcolor[HTML]{FFCCCB}\textbf{0.0894} & \cellcolor[HTML]{FFCCCB}\textbf{0.0008} & \cellcolor[HTML]{D4F1F9}0.0895 &  \cellcolor[HTML]{D4F1F9}0.0009   & 0.0928  & 0.0010   & 0.0974  & 0.0011   & 0.1185 & 0.0013 & 0.1289 & 0.0015  \\ 

\textbf{S1}  & \cellcolor[HTML]{FFCCCB}\textbf{0.0948} & \cellcolor[HTML]{FFCCCB}\textbf{0.0008} & \cellcolor[HTML]{D4F1F9}0.0963 &  \cellcolor[HTML]{D4F1F9}0.0010   & 0.1005  &  \cellcolor[HTML]{D4F1F9}0.0010   & 0.1201  & 0.0014   & 0.1196  & 0.0013 & 0.1335 &  0.0016  \\ 
\midrule
\textbf{Ave.} & \cellcolor[HTML]{FFCCCB}\textbf{0.0902} & \cellcolor[HTML]{FFCCCB}\textbf{0.0008} & \cellcolor[HTML]{D4F1F9}0.0909 & \cellcolor[HTML]{D4F1F9}0.0009 & 0.0932 & 0.0010 & 0.1200 & 0.0013  & 0.1184 &0.0013  & 0.1327 & 0.0015 \\
\bottomrule
\end{tabular}
\end{table*}

\subsection{Performance Metric}

\textbf{Communication reliability}:
To evaluate the accuracy among different demodulators, we measure the bit error rate (BER) per sample as the primary metric, which is defined as
\vspace{-0.2cm}
\begin{equation}
     \text{BER}=\frac{\sum_{u=1}^{U}\sum_{n=1}^{N_{t_{u}}}\vert \hat{\bm b}(u,n)-{\bm b}(u,n)\vert}{\vert \mathcal{X}\vert U\sum_{u=1}^{U}N_{t_{u}}},
     \vspace{-0.2cm}
\end{equation}
where $\bm b(u,n)$ is the transmitted bit sequence of the $n$-th symbol from the $u$-th user, and $\hat{\bm b}(u,n)$ is its estimation by the demodulators.

\textbf{Communication efficiency}:
We measure the throughput as the metric to evaluate the transmission data rate achieved by these demodulators. The throughput is defined as the number of successfully delivered bits per unit time, which can be calculated as
\vspace{-0.2cm}
\begin{equation}
\text{Throughput} \!=\! \frac{ \sum_{u=1}^{U} \!\sum_{n=1}^{N_{t_{u}}} \left( \!\vert \mathcal{X}\vert \! - \!\vert \hat{\bm b}(u,n)\! -\! {\bm b}(u,n) \vert \! \right) }{ T_{\text{total}} },
\vspace{-0.2cm}
\end{equation}
where $L_{b}$ is the number of bits per symbol, and $T_{total}$ is the total time duration for transmitting the considered symbols.

\textbf{Computational complexity}:
In addition to the communication reliability and efficiency, we also assess the computational complexity using floating point operations (FLOPs) and inference latency (ms per sample).

%%%%%% hardware configuration

\subsection{Comparative Study}

We first evaluate the full-shot demodulation accuracy of WiFo-MUD-Base under an SNR of 0 dB. After pre-training on all training datasets, WiFo-MUD is tested on validation sets following the same distribution and configurations. In contrast, we trained and evaluated separate models for ALD on channel data of identical dimensions, and independently trained CoDiPhy and OAMP-Net on each distinct configuration dataset.
To assess performance in practical systems, we also simulated a transmission system incorporating channel coding and decoding, specifically using LDPC codes with a rate of 0.5. Table~\ref{tab:full_shot} summarizes the full-shot accuracy of each method, both with and without channel coding. The results show that the proposed WiFo-MUD achieves the lowest BER across all datasets under both conditions, demonstrating its superiority. The diffusion-based ALD approach ranks second on most datasets, significantly outperforming the linear demodulator and OAMP-Net, which confirms the advantage of iterative sampling under complex channel distributions.

We further compare the zero-shot performance of each method, i.e., the performance of pre-trained models on out-of-distribution (OOD) data, as shown in Table~\ref{tab:zero_shot}. Tests are conducted across diverse scenarios including both simulations and real-world measurements. The results reveal a decline in accuracy for all learning-based models on OOD data, with OAMP-Net exhibiting the most significant drop, underscoring the challenge of distribution shift to the generalization of deep learning models.
In contrast, WiFo-MUD achieves state-of-the-art accuracy on most OOD datasets. This strong generalization is attributed to its precise modeling of noise distributions during pre-training, and the utilization of statistical characteristics during model inference. %Furthermore, when channel coding and decoding are applied, WiFo-MUD maintains a lower BER and preserves its performance advantage over other methods, confirming consistent robustness on both in-distribution and OOD data.
It is also notable that the performance gap between WiFo-MUD and ALD narrows under OOD settings. This phenomenon can be explained by the design of ALD-style methods, which only require generalization to channel dimensions while keeping other computations model-agnostic. This property grants ALD improved adaptability to certain types of distribution shifts. %Despite this, WiFo-MUD's architecture and training strategy still provide it with stronger overall generalization capabilities.

In Fig.~\ref{fig:ber}, we compare the averaged bit error rate (BER) of different demodulators as the variation of SNR under a system configuration with $N_t=2$, $U=4$, $N_r=16$ and QPSK modulation. At low SNR regimes, learning-based demodulators generally achieve lower BER than LMMSE demodulator. However, some of them, such as OAMP-Net, are outperformed by LMMSE at high SNR. In contrast, WiFo-MUD consistently achieves the lowest BER across all SNR levels, yielding a gain of nearly 1 dB over ALD. This accuracy advantage is preserved after introducing channel coding, demonstrating the competitive reliability of WiFo-MUD in communications.

Fig.~\ref{fig:throughput} further illustrates the throughput performance of different demodulation schemes with and without LDPC channel coding. The proposed WiFo-MUD exhibits significant throughput gains in both scenarios. Without coding, conventional linear demodulators show limited throughput growth due to weak interference suppression, while WiFo-MUD, by leveraging multi-user alignment and conditional denoising, effectively mitigates inter-user interference and approaches the theoretical throughput upper bound. When LDPC coding is applied, all methods benefit from coding gains, yet WiFo-MUD still outperforms others, notably maintaining higher throughput in low-SNR regions. These results confirm the efficiency of WiFo-MUD approach in various scenarios.

\begin{figure}[t]
  \vspace{-0.1cm}
  \centering
  \includegraphics[width=0.9\linewidth]{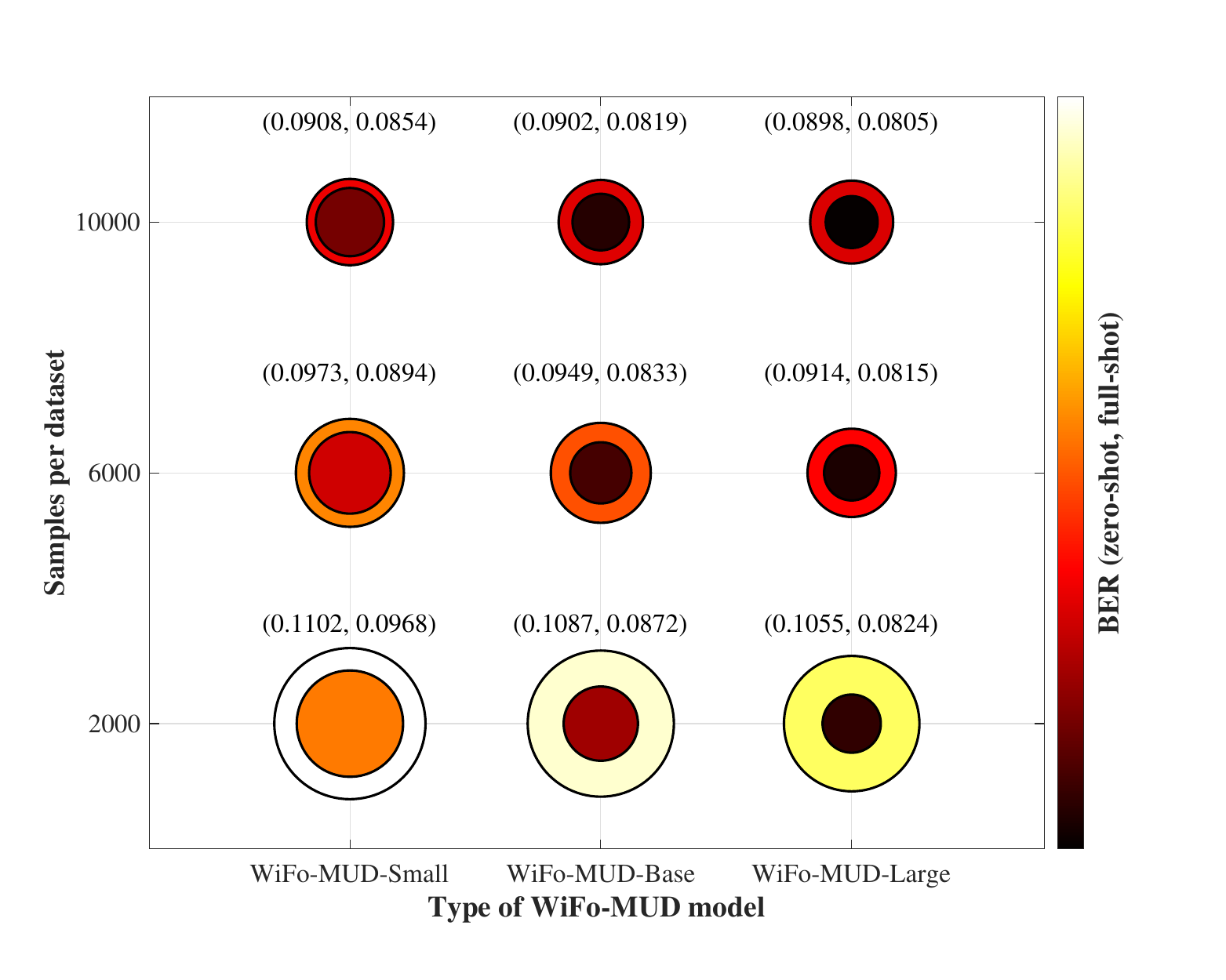}
  \vspace{-0.1cm}
  \captionsetup{font=small}
  \caption{Full-shot and zero-shot BER under varying WiFo-MUD model and dataset scales.}
  \label{fig:scale}
  \vspace{-0.1cm}
\end{figure}

\begin{figure}[t]
  \vspace{-0.0cm}
  \centering
  \includegraphics[width=1.0\linewidth]{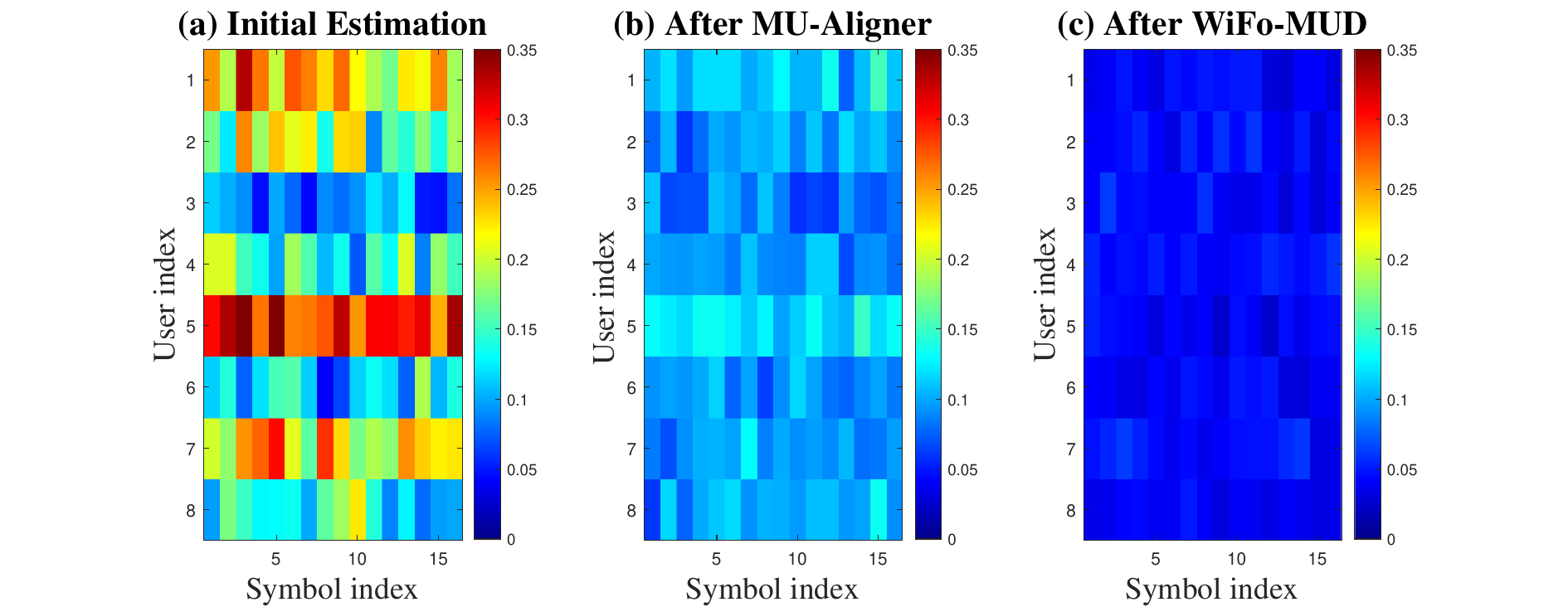}
  \vspace{-0.2cm}
  \captionsetup{font=small}
  \caption{Estimation error level visualization across different processing stages in WiFo-MUD.}
  \label{fig:noise}
  \vspace{-0.5cm}
\end{figure}

We now present a hyperparameter analysis of the proposed method, focusing on the sensitivity to patch size and performance under varying model and dataset scales. By varying the patch size used to segment the channel conditions in the conditional embeddings, we pre-trained and tested the model in dataset Q1 while comparing computational complexity in Table~\ref{tab:patch_size}. Results indicate that excessively small patches allow the model to capture fine-grained correlations within channel conditions, but hinder the extraction of large-scale dependencies, leading to degraded accuracy. Conversely, too-large patches result in too-coarse feature extraction, which also reduces performance. Thus, selecting an appropriate patch size is essential for effective feature extraction. Furthermore, increasing the patch size reduces the number of tokens input to the WirelessDiT blocks, thereby decreasing the overall computational cost.

We further compare the performance of three WiFo-MUD model sizes pre-trained on datasets of different scales in Fig.~\ref{fig:scale}. When trained on datasets of the same size, both full-shot and zero-shot accuracy improve as model parameters increase. The gain in full-shot accuracy is more pronounced, while zero-shot improvement is relatively gradual. For a fixed model size, increasing the pre-training data volume consistently enhances both full-sample and zero-shot accuracy, with the latter benefiting more significantly. This suggests that larger training datasets effectively improve generalization to out-of-distribution data. These observations preliminarily indicate that, similar to large language models, scaling effects also exist in foundational models for wireless communications.

\begin{table}[t]
\centering
\renewcommand\arraystretch{1.3}
\caption{Performance with different patch size}
\label{tab:patch_size}
\begin{tabular}{ccc}
\toprule
\textbf{Patch Size}& \textbf{BER on Q1} & \textbf{FLOPs (M)} \\
\midrule \midrule
(2,   2)     & 0.0812 & 38.45     \\
(2,   4)     &    0.0805    &   35.92        \\
(4,   2)     &  \cellcolor[HTML]{D4F1F9}\underline{0.0794}      &  35.92         \\
(4,   4)     & \cellcolor[HTML]{FFCCCB}\textbf{0.0782} &  32.86   \\
(4,   8)     &   0.0812     &   \cellcolor[HTML]{D4F1F9}\underline{30.67}        \\
(8,   4)     &   0.0820     & \cellcolor[HTML]{D4F1F9}\underline{30.67}          \\
(8,   8)     & 0.0861 &  \cellcolor[HTML]{FFCCCB}\textbf{29.73}     \\
\bottomrule
\end{tabular}
\vspace{-0.00cm}
\end{table}

\begin{table}[t]
\centering
\renewcommand\arraystretch{1.3}
\caption{Ablation on the MU-Aligner module}
\label{tab:mu_aligner}
\begin{tabular}{@{} p{2.3cm} p{2.1cm} p{2.1cm}  @{}}
\toprule
Model  & w/ MU-Aligner      & w/o MU-Aligner    \\ \midrule  \midrule
Full-shot BER       & \textbf{0.0782} & 0.0881    \\ 
Zero-shot BER         & \textbf{0.0891} & 0.0974   \\ 
Inference Time [ms]&   0.57   & 0.53   \\ 
\bottomrule
\end{tabular}
\vspace{-0.1cm}
\end{table}

\begin{table}[t]
\centering
\renewcommand\arraystretch{1.3}
\caption{Ablation on the distillation scheme}
\label{tab:CD}
\begin{tabular}{@{} p{2.3cm} p{0.66cm} p{0.66cm} p{0.66cm} p{0.66cm} p{0.66cm} p{0.66cm} @{}}
\toprule
Inference steps  & 1      & 2     & 5      & 10      & 100      & 1000   \\ \midrule  \midrule
Full-shot BER             & 0.0819 & 0.0816 & 0.0814 & 0.0812  & \cellcolor[HTML]{D4F1F9}\underline{0.0810}  & \cellcolor[HTML]{FFCCCB}\textbf{0.0809}   \\ 
Zero-shot BER             & 0.0902 & 0.0899 & 0.0897 & 0.0894  & \cellcolor[HTML]{D4F1F9}\underline{0.0893}   & \cellcolor[HTML]{FFCCCB}\textbf{0.0892}  \\ 
Inference Time [ms]& \cellcolor[HTML]{FFCCCB}\textbf{0.59}   & \cellcolor[HTML]{D4F1F9}\underline{1.08}   & 2.46   & 5.49    & 58.96    & 602.14    \\ 
\bottomrule
\end{tabular}
\end{table}

\begin{table}[t]
\centering
\renewcommand\arraystretch{1.3}
\caption{Ablation on the distillation loss design}
\label{tab:loss}
\begin{tabular}{@{} p{3.0cm} p{0.8cm} p{0.8cm} p{0.8cm} p{1.4cm}  @{}}
\toprule
Consistency loss weight  & 0      & 1     & 0.5      & adaptive    \\ \midrule  \midrule
Full-shot BER             & 0.0826 & 0.0823 & \cellcolor[HTML]{D4F1F9}\underline{0.0820} & \cellcolor[HTML]{FFCCCB}\textbf{0.0819}   \\ 
Zero-shot BER             & 0.0941 & 0.0907 & \cellcolor[HTML]{D4F1F9}\underline{0.0906} & \cellcolor[HTML]{FFCCCB}\textbf{0.0902}  \\ 
\bottomrule
\end{tabular}
\vspace{-0.35cm}
\end{table}

\begin{table}[t]
\centering
\renewcommand\arraystretch{1.3}
\caption{Ablation on the user grouping scheme}
\label{tab:user_group}
\begin{tabular}{ccccc}
\toprule
Grouping threshold     & $N_{r}$ & $N_{r}/2$ & $N_{r}/4$ & ALD    \\ 
\midrule \midrule
Full-shot BER           & 0.0833   & \cellcolor[HTML]{D4F1F9}\underline{0.0819}     & \cellcolor[HTML]{FFCCCB}\textbf{0.0815}     & 0.0843 \\ 
Zero-shot BER           & 0.0913   & \cellcolor[HTML]{D4F1F9}\underline{0.0902}    & \cellcolor[HTML]{FFCCCB}\textbf{0.0898}     & 0.0909 \\ 
Inference Time {[}ms{]} & \cellcolor[HTML]{FFCCCB}\textbf{0.33}    & \cellcolor[HTML]{D4F1F9}\underline{0.59}      & 1.26       & 40.46  \\ 
\bottomrule
\end{tabular}
\vspace{-0.35cm}
\end{table}

\begin{figure}[t]
  \vspace{-0.1cm}
  \centering
  \includegraphics[width=0.93\linewidth]{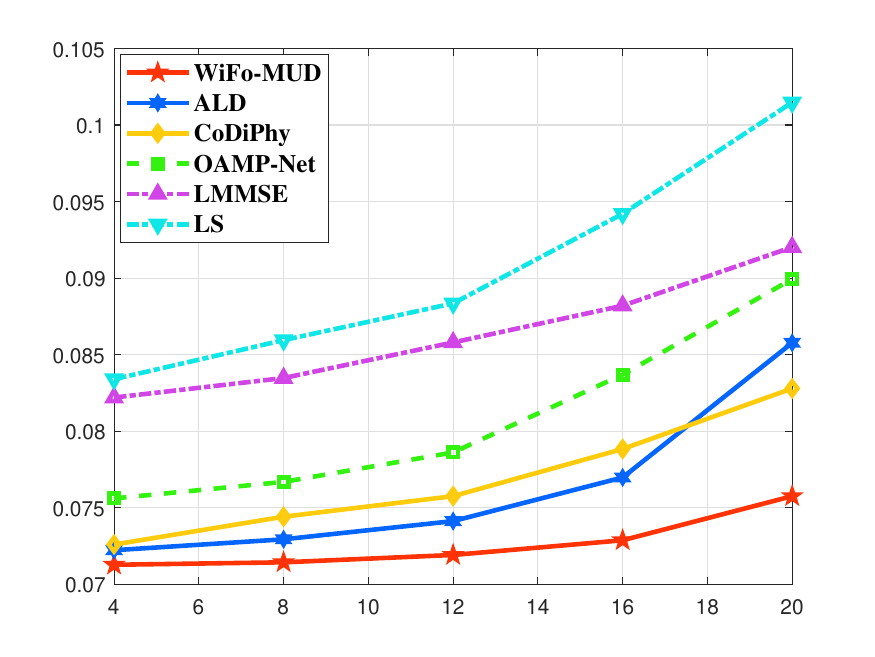}
  \vspace{-0.25cm}
  \captionsetup{font=small}
  \caption{BER comparison under varying number of users.}
  \label{fig:more_user}
  \vspace{-0.45cm}
\end{figure}

\begin{table*}[t]
\centering
\renewcommand\arraystretch{1.3}
\caption{Performance and complexity comparison among different schemes}
\label{tab:complexity}
\begin{tabular}{ccccccc}
\toprule
\textbf{Methods}  & \textbf{WiFo-MUD-Small} & \textbf{WiFo-MUD-Base} & \textbf{WiFo-MUD-Large} & \textbf{ALD} & \textbf{CoDiPhy}  & \textbf{OAMP-Net} \\ \midrule
Full-shot SER   & 0.0853    & \cellcolor[HTML]{D4F1F9}\underline{0.0819}   & \cellcolor[HTML]{FFCCCB}\textbf{0.0803}    & 0.0843    & 0.0857    & 0.0879   \\
Zero-shot SER   & 0.0906     & \cellcolor[HTML]{D4F1F9}\underline{0.0902}   & \cellcolor[HTML]{FFCCCB}\textbf{0.0895}    & 0.0909    & 0.0932    & 0.1200   \\
Parameters (M)  & 1.29       & 3.92     & 5.83      & 1.34      & \cellcolor[HTML]{D4F1F9}\underline{1.19}      & \cellcolor[HTML]{FFCCCB}\textbf{0.56}     \\
FLOPs (M)       & \cellcolor[HTML]{D4F1F9}\underline{15.47}      & 32.86    & 59.35     & 15.80*70  & 5.16*500  & \cellcolor[HTML]{FFCCCB}\textbf{12.64}    \\
Inference Time (ms) & \cellcolor[HTML]{D4F1F9}\underline{0.36}   & 0.59     & 0.73      & 40.46     & 34.61     & \cellcolor[HTML]{FFCCCB}\textbf{0.34}     \\ 
\bottomrule
\end{tabular}
\end{table*}

We further validate the effectiveness of MU-Aligner in mitigating the multi-user imbalance issue in Fig.~\ref{fig:noise} and Table \ref{tab:mu_aligner}. Fig.~\ref{fig:noise} visualizes the noise levels of the signal estimates for 8 users from dataset Q1, comparing the initial LMMSE coarse estimates, the estimates after applying the MU-Aligner, and the final estimates after the complete WiFo-MUD processing. The comparison demonstrates that MU-Aligner effectively equalizes the SNRs across users through its timestep-aware denoising. Subsequently, the WirelessDiT further reduces the overall estimation error. An ablation study on the MU-Aligner module is presented in Table~\ref{tab:mu_aligner}. The results show that the module achieves a significant improvement in final demodulation accuracy with only a marginal increase in latency.

We further conduct ablation experiments to evaluate the effectiveness of the proposed communication-aware consistency distillation mechanism. For the pre-trained and distilled WiFo-MUD-Base model, we maintained the user grouping threshold while varying the number of inference steps. The detailed inference procedure is described in \cite{CM}. As shown in Table~\ref{tab:CD}, both full-shot and zero-shot accuracy improve as the number of inference steps increases, owing to enhanced denoising capability. Compared to the original 1000-step inference, single-step inference exhibits only a marginal decline in accuracy, confirming the effectiveness of our distillation in preserving knowledge of the noise distribution. Meanwhile, the total inference time increases approximately linearly with the number of steps. Thus, by appropriately selecting the number of inference steps, the model can maintain its accuracy advantage without introducing additional latency.

We also investigate the impact of different loss functions during the distillation process in Table~\ref{tab:loss}. The variation in loss weighting reflects the trade-off between consistency loss and MSE loss. It can be observed that employing only the MSE loss hinders effective knowledge transfer from the teacher model, leading to severely degraded zero-shot performance. Conversely, relying solely on consistency loss results in non-negligible accuracy degradation compared to the teacher model. Although the combined loss with fixed weights improves over single-loss configurations, it still causes a drop in zero-shot accuracy. In contrast, our proposed adaptive weighting strategy achieves the best performance in both full-shot and zero-shot scenarios.

We also validate the effectiveness of the dynamic user grouping strategy. Table~\ref{tab:user_group} compares the demodulation accuracy of WiFo-MUD under different upper limits on the number of data streams. The theoretical maximum is set by $N_{r}$. A lower limit implies fewer users processed per group and thus a larger number of groups. Results show that demodulation accuracy improves as the group size decreases, indicating that finer-grained grouping alleviates inter-user interference during inference and reduces the overall BER. However, this comes at the cost of an increased number of inference runs and higher computational load. This behavior is analogous to the mechanism of non-linear decision-feedback demodulators.

In Fig.~\ref{fig:more_user}, we analyze the accuracy of different schemes in scenarios with varying user access density. With $N_r=32$ and each user equipped with 2 antennas, the number of users increases from 4 to 20. Under such conditions, the BER of all baseline demodulators rises due to growing SNR disparities and inter-user interference. Among them, the ALD demodulator suffers from error accumulation in the score network under high user density, which is further amplified during iterative sampling, resulting in lower accuracy than CoDiPhy. In contrast, the dynamic user grouping strategy in WiFo-MUD effectively mitigates inter-user interference with only marginal complexity overhead, enabling it to maintain high accuracy as the number of users increases.

Finally, we compare the computational overhead of each method. Although WiFo-MUD contains more parameters than other deep learning-based models, it requires only single-step inference after consistency distillation, resulting in significantly fewer FLOPs than iterative diffusion-based benchmarks. The MU-Aligner, designed with minimal layers and parallel inference support, contributes negligible latency. As summarized in Table~\ref{tab:complexity}, the inference time of WiFo-MUD is comparable to that of small-model approaches, confirming its suitability for real-time systems.

\section{Conclusion}

This paper introduced WiFo-MUD, a foundation model for multi-user signal demodulation under heterogeneous configurations. The architecture, comprising MU-Aligner and WirelessDiT modules, was tailored for wireless signal characteristics and multi-user demodulation tasks. Through diffusion-based self-supervised pre-training, the model recovered transmitted signals from coarse and imbalanced estimates. To enhance inference, a communication-aware consistency distillation scheme enabled single-step sampling, while a dynamic user-grouping strategy was implemented. Evaluated on large-scale heterogeneous datasets, WiFo-MUD achieved state-of-the-art performance in both full-shot and zero-shot settings with a single model, demonstrating strong potential for physical-layer communication systems.

%\appendices

% you can choose not to have a title for an appendix
% if you want by leaving the argument blank

% use section* for acknowledgment
%\section*{Acknowledgment}

%The authors would like to thank...

% Can use something like this to put references on a page
% by themselves when using endfloat and the captionsoff option.
\ifCLASSOPTIONcaptionsoff
  \newpage
\fi

\bibliographystyle{IEEEtran}
\bibliography{IEEEabrv,myrefs}

\end{document}